\documentclass[aps,pre,groupedaddress,showpacs,amsmath,amssymb,twocolumn]{revtex4-2}

\usepackage{graphicx,color}
\usepackage{epstopdf}
\usepackage{hyperref}
\usepackage{makecell}
\usepackage{bm}
\newcommand{\hsfn}[1]{{\textcolor{black}{#1}}}
\newcommand{\hs}[1]{{\textcolor{black}{#1}}}
\newcommand{\hsn}[1]{{\textcolor{black}{#1}}}

\begin{document}

\def\ldotsplus{\mathinner{\ldotp\ldotp\ldotp\ldotp}}
\def\fourdots{\relax\ifmmode\ldotsplus\else$\m@th \ldotsplus\,$\fi}


\title{Properties of the dissipation functions for passive and active systems}
\author{Harsh Soni}
\affiliation{School of Physical Sciences, IIT Mandi, Kamand, Mandi, HP 175005 India}
\date{\today}
\begin{abstract}
The dissipation function for a system is defined as the natural logarithm of the ratio between probabilities of a \hsn{trajectory} and its time-reversed trajectory, and its probability distribution follows a well-known relation called the \hsn{fluctuation} theorem. Using the generic Langevin equations, we derive the expressions of the dissipation function for passive and active systems. For passive systems, 
the dissipation function depends only on the initial and the final values of the dynamical variables of the system, not on the trajectory of the system. 
Furthermore, it does not depend explicitly on the reactive or dissipative coupling coefficients of the generic Langevin equations. In addition, we study a 1D case numerically to verify the fluctuation theorem with the form of the dissipation function we obtained. 
For active systems, we define the work done by active forces along a trajectory. If the probability distribution of the dynamical variables is symmetric under time reversal, in both cases, the \hsn{average rate} of change of the dissipation function with trajectory duration is nothing but the average entropy production rate of the system and reservoir.
\end{abstract}



\maketitle
\section{Introduction}
Irreversibility of a system can be quantified by using the dissipation function which is defined as the natural logarithm of the ratio of the probability density of a trajectory to that of its time-reversed trajectory. The probability distribution function of the dissipation function exhibits an interesting symmetry relation known as the fluctuation theorem~\cite{evans_first_prl,evans_ad_phy,Searles2004}. 
The fluctuation theorem has been substantially explored using theory~\cite{Jarz_equlity_scince_2002,Seifert_prl_2005,seifert2006prl,seifert2012stochastic,gallavotti_epjb_2008,jarzynski2007comparison,ifr_pnas_2011, AFR_EPL_2014, soni2019flocks} and experiment~\cite{op_tweez_exp_prl_2002,crook_dna_nature_2005,op_tweez_exp_prl_2004,evans1994pre,khan2010out,kink_nitin_prl_2011,nk_hs_sr_as2014}. 
For stochastic processes, it has been investigated mainly for the single-particle or single-variable case~\cite{Seifert_prl_2005,seifert2006prl,aron2016dynamical}.
Moreover, \hsn{little} attention has been paid to the systems described by the Langevin equations with multiplicative noise, except \hsn{for} a few studies~\cite{aron2016dynamical,soni2019flocks}.
%


This paper discusses the fluctuation relations for a wide class of systems described by the generic Langevin equations~\cite{dengler2015,mazenkobook}.
Assuming that the slow variables of a system vary much slower than its microscopic degrees of \hsn{freedom, one} can consider that the system is always in local thermodynamic equilibrium at temperature $T$. The dynamics of such systems is well-explained by the generic Langevin equations. We consider the active as well as passive systems.  
We use the path integral approach to calculate the probability density of a trajectory of the system with $\alpha$-discritization~\cite{lau2007state,aron2016dynamical,cugliandolo2017rules}. 

Our main results are as follows.
\hsn{We first show that the generic Langevin equations describe a passive system, irrespective of the  value of $\alpha$. }
We then derive the expression of the dissipation function for passive systems relaxing towards thermodynamic equilibrium. Interestingly,  the dissipation function  is independent of the trajectory followed by the system; it only depends on the initial and the final values of the dynamical variables of the system. Moreover, it is not an explicit function of the coefficients appearing in the generic Langevin equation. \hsn{Using Brownian dynamics simulation, we also verify the fluctuation theorem for a 1D single-particle problem with state-dependent diffusion.} Finally, we construct an expression of the dissipation function for the active systems, and we define the work done by the active forces. For both active and passive systems, the \hsn{average rate} of change of the dissipation function with the time duration is \hsn{the same as} the rate of \hsn{change of} the entropy of the system and reservoir, assuming that the probability distribution of the dynamical variables is invariant under time reversal.

In the next section, we will discuss passive systems, and in section~\ref{actdiss}, we will explore active systems.

\section{Passive systems}
This section is arranged as follows. In the next subsection, we summarize the \hsfn{generic} Langevin equations. We then calculate the ratio of the probabilities of a trajectory and its time-reversed trajectory (see subsection~\ref{ratio}). In subsection~\ref{ftrsub}, fluctuation relations and the dissipation function for the passive systems are presented. In subsection~\ref{qunchsys}, we talk about the quenched systems, along with an example of a system of a single colloidal particle.
\subsection{Generic Langevin equations}\label{glesubs}
\hsn{Here we consider a passive system relaxing towards equilibrium. Its} macroscopic dynamics is described by a set of $n$ number of slow variables $\bm{A}\equiv\{A_1,A_2,....A_n\}$. Let $A_i\to s_i A_i$ under time reversal, where $s_i=1$ and $s_i=-1$ if $A_i$ is even and odd under time reversal, respectively; e.g., $s_i=1$ for position  and $s_i=-1$ for momentum. The generic Langevin equations for the system at temperature $T$ can be written in the following form~\cite{dengler2015,mazenkobook}: 
\begin{equation}\label{gle}
\dfrac{dA_i}{dt}=-\Gamma_{ij}\dfrac{\partial \mathcal{H}}{\partial A_j}+k_\text{B}T\dfrac{\partial \Gamma_{ij}}{\partial A_j}+\eta_i(t).
\end{equation}
where $\mathcal{H}$ is the coarse-grained or effective Hamiltonian of the system and the coefficients $\Gamma_{ij}$ satisfy the following property:
\begin{equation}\label{gamma}
\Gamma_{ij}=s_is_j\Gamma_{ji}.
\end{equation}
In Eq.~\eqref{gle}, the terms with $s_is_j=-1$ are the Poisson bracket or reactive terms, whereas the terms with $s_is_j=1$ are the dissipative terms~\cite{Lubensky_book}.
The last term $\eta_i(t)$ represents the rapid fluctuations due to the dynamics of the microscopic degrees of freedom of the system\hsn{. We assume that $\eta_i(t)$ is white Gaussian noise, and} its autocorrelation function is given by 
\begin{equation}\label{etacor}
\left\langle \eta_i(t)\eta_j(t')\right\rangle =2 k_\text{B}T \Gamma^\text{s}_{ij}\delta(t-t'),
\end{equation}
where $\Gamma^\text{s}_{ij}\equiv(\Gamma_{ij}+\Gamma_{ji})/2$ is the symmetric part of $\Gamma_{ij}$. 
\hsn{From Eq.\eqref{gamma}, $\Gamma^\text{s}_{ij}$ shows the following symmetry property:
\begin{equation}\label{Gammaspro}
\Gamma^\text{s}_{ij}=s_i s_j\Gamma^\text{s}_{ij}.
\end{equation}
Further, it is assumed to be invertible. Here in Eq.~\eqref{gle}, we use Einstein notation, which will be carried through the rest of the paper, unless otherwise stated.}
Writing $\eta_i(t)$ as the linear combination of \hsn{time series of the white Gaussian} noise $\xi_j(t)$ having no correlation with each other \textit{i.e.} $\left\langle\xi_i(t)\xi_j(t)\right\rangle =\delta_{ij}\delta(t-t')$:
\begin{equation}\label{eta}
\eta_i(t)=N_{ij}\xi_j(t),
\end{equation}
where, from Eq.~\eqref{etacor}, $N_{ij}$ is given by the solution of the equations:
\begin{equation}\label{nn}
N_{ik}N_{jk}=2 k_\text{B}T \Gamma^\text{s}_{ij}.
\end{equation}
Since $N_{ik}$ must be real, $\Gamma^\text{s}_{ij}$ must have positive eigenvalues~\cite{Onsager1931}. \hsn{ As $\Gamma^\text{s}_{ij}$ is considered to be invertible, $N_{ik}$ is invertible as well. It should be noted that $N_{jk}$ is not uniquely defined by the above equation. However, $N_{jk}$ is just a dummy matrix which does not appear anywhere in the final results.}
Substituting~\eqref{eta} into  Eq.~\eqref{gle}:
 \begin{equation}\label{gle1}
\dfrac{dA_i}{dt}=-\Gamma_{ij}\dfrac{\partial \mathcal{H}}{\partial A_j}+k_\text{B}T\dfrac{\partial \Gamma_{ij}}{\partial A_j}+N_{ij}\xi_j(t).
\end{equation}
The above stochastic equations have no ambiguity when $N_{ij}$ does not depend explicitly on $\bm{A}$. However, $N_{ij}$ is the function of $\bm{A}$ for many systems; in such cases, the above equations are not well-defined unless their discrete scheme is specified. We here use $\alpha$-discretization method to discretize the above equations~\cite{aron2016dynamical,lau2007state,cugliandolo2017rules}, which \hsn{leads to} a drift of $\alpha N_{lj}  \partial N_{ij}/\partial A_l$ to the value of $A_i$ due to the noise term~\cite{tcgard, cugliandolo2017rules}. On the contrary, the noise terms in the generic Langevin equations represent the thermal fluctuations and do not contribute to the average dynamics of the slow variables. One can eliminate the noise-induced drift by adding a correction term $-\alpha N_{lj}  \partial N_{ij}/\partial A_l$ to Eq.~\eqref{gle1}. Therefore, the generic Langevin equations can be completely described as follows:
\begin{eqnarray}\label{gle2}
\dfrac{dA_i}{dt}&=&-\Gamma_{ij}\dfrac{\partial \mathcal{H}}{\partial A_j}+k_\text{B}T\dfrac{\partial \Gamma_{ij}}{\partial A_j}-\alpha N_{lj}  \dfrac{\partial N_{ij}}{\partial A_l}+N_{ij}\xi_j\\
&\equiv&\mathcal{F}_i+N_{ij}\xi_j
\end{eqnarray}
 with \hsn{their} discrete form
\begin{equation}\label{geldisc}
dA_i(l)=\epsilon\mathcal{F}_i(\bar{\bm{A}}^\text{f}_l)+\sqrt{\epsilon}N_{ij}(\bar{\bm{A}}^\text{f}_l)\xi_j^{l},
\end{equation}
where  $\epsilon$ is the time step, $dA_i(l)\equiv A_i(\epsilon l)-A_i(\epsilon (l-1))$, $\bar{\bm{A}}^\text{f}_l\equiv \alpha \bm{A}(\epsilon l)+(1-\alpha) \bm{A}(\epsilon (l-1))$,
\begin{equation}\label{key}
\mathcal{F}_i\equiv-\Gamma_{ij}\dfrac{\partial \mathcal{H}}{\partial A_j}+k_\text{B}T\dfrac{\partial \Gamma_{ij}}{\partial A_j}-\alpha N_{lj}  \dfrac{\partial N_{ij}}{\partial A_l},
\end{equation}
and,
\begin{equation}\label{key}
\xi_j^l\equiv \dfrac{1}{\sqrt{\epsilon}}\int^{l\epsilon}_{(l-1)\epsilon}\xi_j(t)dt.
\end{equation}
\hsn{ are the series} of random numbers having normal distribution  with standard deviation 1 and mean 0. The parameter $\alpha$ can take any \hsn{``absolute constant''} between 0 and 1; $\alpha=0$ and $\alpha=1/2$ cases are referred to as It\^{o} and Stratonovich methods, respectively. \hsn{As we have another parameter $\alpha$ in the problem now, one of the questions we ask here is, do different values of $\alpha$ correspond to different systems? If yes, do all the values of $\alpha$ belong to passive systems? }

Based on the behavior under time reversal, dividing  $\mathcal{F}_i$ into the following \hsn{three parts $\mathcal{F}^\text{s}_i$, $\mathcal{F}^\text{a}_i$ and $\mathcal{F}^\text{N}_i$: 
\begin{eqnarray}\label{F}
&&\mathcal{F}^\text{s}_i(\bm{A})=-\Gamma^\text{s}_{ij}\dfrac{\partial \mathcal{H}}{\partial A_j}+k_\text{B}T\dfrac{\partial \Gamma^\text{s}_{ij}}{\partial A_j}\\
&&\mathcal{F}^a_i(\bm{A})=-\Gamma^a_{ij}\dfrac{\partial \mathcal{H}}{\partial A_j}+k_\text{B}T\dfrac{\partial \Gamma^\text{a}_{ij}}{\partial A_j}\\
\label{FN}
&&\mathcal{F}^\text{N}_i=- \alpha N_{kj}\dfrac{\partial N_{ij}}{\partial A_k}
\end{eqnarray}
where} $\Gamma^\text{a}_{ij}\equiv(\Gamma_{ij}-\Gamma_{ji})/2$ is the antisymmetric part of $\Gamma_{ij}$. From Eq.\eqref{gamma}\hsn{, $\Gamma^\text{a}_{ij}$ exhibits the following symmetry property (not in Einstein notation):
\begin{equation}\label{Gammaapro}
\Gamma^\text{a}_{ij}=-s_is_j\Gamma^\text{a}_{ij}.
\end{equation}
Since} $\mathcal{H}(\bm{s}\circ\bm{A})=\mathcal{H}(\bm{A})$ and $\Gamma^\text{s}_{ij}(\bm{s}\circ\bm{A})=\Gamma^\text{s}_{ij}(\bm{A})$~\cite{dengler2015}, from Eqs.~\eqref{Gammaspro} \&~\eqref{Gammaapro}, under time reversal,
\hsn{\begin{eqnarray}
\label{Fprop1}
&&\bm{\mathcal{F}}^\text{s}(\bm{A})\to\bm{\mathcal{F}}^\text{s}(\bm{s}\circ\bm{A})=\bm{s}\circ \bm{\mathcal{F}}^\text{s}(\bm{A}),\\
\label{Fprop2}
&&\bm{\mathcal{F}}^\text{a}(\bm{A})\to\bm{\mathcal{F}}^\text{a}(\bm{s}\circ\bm{A})=-\bm{s}\circ \bm{\mathcal{F}}^\text{a}(\bm{A}).
\end{eqnarray}}
where $\circ$ stands for Hadamard product \textit{i.e.} $\bm{s}\circ\bm{A}\equiv\{s_1A_1,s_2A_2,....s_nA_n\}$. \hsn{For given $\Gamma^\text{s}_{ij}$, $\mathcal{F}^\text{s}_i(\bm{A})$ and $\mathcal{F}^\text{a}_i(\bm{A})$ do not depend on $N_{ij}$. In general, $\mathcal{F}^\text{N}_i$ does not follow any of the above time reversal symmetries.}

\subsection{The ratio between the  probability densities of a trajectory and its time-reversed trajectory}~\label{ratio}
Let $p_0(\bm{A})$ be the probability distribution function of $\bm{A}$ at $t=0$.
\hsn{In $\epsilon\to 0$ limit, the} probability density of a trajectory of the system $(\bm{A}_0,\bm{A}_1,\bm{A}_2,.....\bm{A}_N)$ (here $\bm{A}_l\equiv\bm{A}(l\epsilon)$) between $t=0$ and $t=\tau \equiv N \epsilon$ is given by~\cite{soni2019flocks} (see Appendix~\ref{Appdf})
\hsn{\begin{widetext}
\begin{eqnarray}\label{Pform}
\nonumber
P&\simeq&p_0(\bm{A}_0)\prod^{N}_{l=1}\left\lbrace    \dfrac{(2k_\text{B}T)^{-1/2}}{(2\pi \epsilon)^{n/2}}\exp\Biggl[-\dfrac{1}{4\epsilon k_\text{B}T} \left[dA_i(l)-\epsilon\mathcal{F}^\text{s}_i(\bar{\bm{A}}^\text{f}_l)-\epsilon\mathcal{F}^\text{a}_i(\bar{\bm{A}}^\text{f}_l)\right] {\hsn{(\Gamma^{s}}^{-1})_{ij}}(\bar{\bm{A}}^\text{f}_l)\left[dA_j(l)-\epsilon\mathcal{F}^\text{s}_j(\bar{\bm{A}}^\text{f}_l)-\epsilon\mathcal{F}^\text{a}_j(\bar{\bm{A}}^\text{f}_l)\right]\Biggr. \right. \\
\nonumber
&&\left. -\alpha \left[ \left( dA_i(l)-\epsilon\mathcal{F}^\text{s}_i(\bar{\bm{A}}^\text{f}_l)-\epsilon\mathcal{F}^\text{a}_i(\bar{\bm{A}}^\text{f}_l)\right)\left[  ({\Gamma^{s}}^{-1})_{ij}\dfrac{\partial \Gamma^{s}_{jk}}{\partial A_k}\right]_{\bar{\bm{A}}^\text{f}_l}+\epsilon\left[\dfrac{\partial \mathcal{F}^\text{s}_i}{\partial A_i}+\dfrac{\partial \mathcal{F}^\text{a}_i}{\partial A_i}\right]_{\bar{\bm{A}}^\text{f}_l}\right]\right]
 \text{det}\left(  \bm{\Gamma^\text{s}}(\bar{\bm{A}}^\text{f}_{l})\right)^{-1/2}  \\
&&\times\left.   \exp \left[\alpha^2\epsilon k_\text{B}T\left[\dfrac{\partial^2 \Gamma^{s}_{ij}}{\partial A_i\partial A_j} -\dfrac{\partial \Gamma^{s}_{ik}}{\partial A_k}({\Gamma^{s}}^{-1})_{ij} \dfrac{\partial \Gamma^{s}_{jp}}{\partial A_p}\right]_{\bar{\bm{A}}^\text{f}_l}\right]\right\rbrace.
\end{eqnarray}
\end{widetext}
The $\epsilon^{3/2}$- and higher-order terms have been neglected here. It is apparent from the above expression that, for given $\Gamma^{\text{s}}_{ij}$, $P$ is independent of $N_{ij}$. So no statistical property of the system has a dependence upon the choice of $N_{ik}$. Therefore, as mentioned earlier, $N_{ik}$ is merely a dummy matrix. The probability density $P$ depends on $\alpha$; thus, the different values of $\alpha$ correspond to different systems. Later in this subsection, we will see that Eq.~\eqref{gle2} provides the dynamics of a passive system for any $\alpha$.}
The time-reversed trajectory of the trajectory $(\bm{A}_0,\bm{A}_1,\bm{A}_2,.....\bm{A}_N)$  would be $(\bm{s}\circ\bm{A}_N,\bm{s}\circ\bm{A}_{N-1},.....\bm{s}\circ\bm{A}_1)$, so its probability density can be calculated by replacing $\bm{A}_i$ by $\bm{s}\circ\bm{A}_{N-i}$ in the above equation\hsn{; that is, (see Appendix~\ref{revp})
\begin{widetext}
\begin{eqnarray}\label{Prform}
\nonumber
P_r&\simeq&\prod^{N}_{l=1}\left\lbrace    \dfrac{(2k_\text{B}T)^{-1/2}}{(2\pi \epsilon)^{n/2}}\exp\Biggl[-\dfrac{1}{4\epsilon k_\text{B}T} \left[-dA_i(l)-\epsilon\mathcal{F}^\text{s}_i(\bar{\bm{A}}^\text{r}_{l})+\epsilon\mathcal{F}^\text{a}_i(\bar{\bm{A}}^\text{r}_{l})\right] {\hsn{(\Gamma^{s}}^{-1})_{ij}}(\bar{\bm{A}}^\text{r}_{l})\left[-dA_j(l)-\epsilon\mathcal{F}^\text{s}_j(\bar{\bm{A}}^\text{r}_{l})+\epsilon\mathcal{F}^\text{a}_j(\bar{\bm{A}}^\text{r}_{l})\right]\Biggr. \right. \\
\nonumber
&&\left. -\alpha \left[ \left( -dA_i(l)-\epsilon\mathcal{F}^\text{s}_i(\bar{\bm{A}}^\text{r}_{l})+\epsilon\mathcal{F}^\text{a}_i(\bar{\bm{A}}^\text{r}_{l})\right)\left[  ({\Gamma^{s}}^{-1})_{ij}\dfrac{\partial \Gamma^{s}_{jk}}{\partial A_k}\right]_{\bar{\bm{A}}^\text{r}_{l}}+\epsilon\left[\dfrac{\partial \mathcal{F}^\text{s}_i}{\partial A_i}-\dfrac{\partial \mathcal{F}^\text{a}_i}{\partial A_i}\right]_{\bar{\bm{A}}^\text{r}_{l}}\right]\right]
 \text{det}\left(  \bm{\Gamma^\text{s}}(\bar{\bm{A}}^\text{r}_{l})\right)^{-1/2}  \\
&&\times\left.   \exp \left[\alpha^2\epsilon k_\text{B}T\left[\dfrac{\partial^2 \Gamma^{s}_{ij}}{\partial A_i\partial A_j} -\dfrac{\partial \Gamma^{s}_{ik}}{\partial A_k}({\Gamma^{s}}^{-1})_{ij} \dfrac{\partial \Gamma^{s}_{jp}}{\partial A_p}\right]_{\bar{\bm{A}}^\text{r}_{l}} \right]\right\rbrace p_0(\bm{s}\circ\bm{A}_N).
\end{eqnarray}
\end{widetext}
 In $\epsilon\to0$ limit, the ratio $P/P_r$ takes the following form (see Appendix~\ref{Pratiosup}):}
\begin{equation}\label{Pratio}
\dfrac{P}{P_r}=\dfrac{p_0(\bm{A}(0))}{p_0(\bm{s}\circ\bm{A}(\tau))}\exp\left[-\dfrac{1}{k_\text{B}T} \left( \mathcal{H}(\bm{A}(\tau))-\mathcal{H}(\bm{A}(0) )\right) \right]
\end{equation}  
\hsn{In the stationary state, $p_0(\bm{A})=\exp(-\mathcal{H}(\bm{A})/k_\text{B}T)/\mathcal{Z}$ (see Appendix~\ref{fps}), the above equation then yields
\begin{equation}\label{Pratio1}
P=P_r.
\end{equation}
It implies that any system whose dynamics is given by Eq.~\eqref{gle2} has the time reversal symmetry in its stationary state, regardless of the value of $\alpha$. Hence, Eq.~\eqref{gle2} describes a passive system for any value of $\alpha$ between 0 and 1.}
\subsection{Fluctuation relations and the dissipation function for the passive systems}~\label{ftrsub}
One can readily show that~\cite{soni2019flocks} the dissipation function $\mathcal{R}_\tau$ for the trajectory $(\bm{A}_0,\bm{A}_1,\bm{A}_2,.....\bm{A}_N)$ defined as 
\begin{equation}\label{key}
\mathcal{R_\tau}=\ln \left[ \dfrac{P}{P_r}\right]
\end{equation}
satisfies the relation
\begin{equation}\label{ft}
\hsn{\dfrac{\mathcal{P}(\mathcal{R}_\tau=X)}{\mathcal{P}(\mathcal{R}_\tau=-X)}=\exp (X),}
\end{equation}
where $\mathcal{P}$ is the probability distribution of $\mathcal{R}_\tau$.
The above relation is known as the fluctuation theorem~\cite{evans_ad_phy,Seifert_prl_2005}. From Eq.~\eqref{Pratio},
\begin{equation}\label{dissf}
\mathcal{R}_\tau=\ln \dfrac{p_0(\bm{A}(0))}{p_0(\bm{s}\circ\bm{A}(\tau))}-\dfrac{1}{k_\text{B}T}\left[  \mathcal{H}(\bm{A}(\tau)) -\mathcal{H}(\bm{A}(0))\right].
\end{equation}
Intriguingly,  $\mathcal{R}_\tau$ does not depend explicitly on $\Gamma_{ij}$. Moreover, it depends only the initial and final values of $\bm{A}$, not on the trajectory followed by $\bm{A}$. Note that the ratio $P/P_r$ in Eq.~\eqref{Pratio} also has the same functional properties.

One can also define the dissipation function for the system as follows~\cite{evans_first_prl}: 
\begin{equation}\label{key}
\mathcal{R}'_\tau=\ln \dfrac{p(\bm{A}(0),\bm{A}(\tau);\tau)}{p(\bm{s}\circ\bm{A}(\tau),\bm{s}\circ\bm{A}(0);\tau)},
\end{equation}
where $p(\bm{A}(0),\bm{A}(\tau);\tau)$ is the net probability that the system goes from $\bm{A}(0)$ to $\bm{A}(\tau)$ in time $\tau$:
\begin{equation}\label{key}
p(\bm{A}(0),\bm{A}(\tau);\tau)=\sum_{} P,
\end{equation}
where the summation is performed over all the trajectories between $\bm{A}(0)$ and $\bm{A}(\tau)$.
Since the ratio $P/P_r$ is independent of the trajectory between $\bm{A}(0)$ and $\bm{A}(\tau)$, $\mathcal{R'}_\tau=\mathcal{R}_\tau$.

The integrated form of the relation~\eqref{ft} is
\begin{equation}\label{ift}
\left\langle \exp(-\mathcal{R}_\tau)\right\rangle=1,
\end{equation}
where angular bracket stands for the ensmeble average.
Using this relation, one can show that~\cite{ft_evan_review,Seifert_prl_2005}
\begin{equation}\label{thermsecc}
\left\langle \mathcal{R}_\tau\right\rangle \geq 0.
\end{equation}
\hsn{In equilibrium, $P=P_r$, thus $\left\langle \mathcal{R}_\tau\right\rangle=\mathcal{R}_\tau=0$. So $\left\langle \mathcal{R}_\tau\right\rangle $} behaves like the change in the entropy of system and it can be used to evaluate that how far the system is from equilibrium. \hsn{Since the solution of the Fokker-Planck equation corresponding to Eq.~\eqref{gle2} does not depend on $\alpha$, $\left\langle \mathcal{R}_\tau\right\rangle $ is constant in  $\alpha$  (see Appendix~\ref{rtaui}). }

Generalizing the expression of the dissipation function given in Eq.~\eqref{dissf} for the trajectories starting at arbitrary time $t$:
\begin{eqnarray}\label{dissft}
\nonumber
\mathcal{R}_\tau(t)&=&\ln \dfrac{p_t(\bm{A}(t))}{p_t(\bm{s}\circ\bm{A}(t+\tau))}\\&&-\dfrac{1}{k_\text{B}T}\left[  \mathcal{H}(\bm{A}(t+\tau)) -\mathcal{H}(\bm{A}(t))\right].
\end{eqnarray}
\hs{The instantaneous irreversibility can be evaluated by calculating $\mathcal{R}_\tau(t)$ in $\tau\to 0$ limit, that is,
\begin{equation}\label{rtau0}
\hsn{\mathcal{R}_\tau(t)\simeq\dot{{s}}(t)\tau+\ln\dfrac{p_t(\bm{A}(t+\tau))}{p_t(\bm{s}\circ\bm{A}(t+\tau))},}
\end{equation}	
where
}
\begin{equation}\label{sdotpass}
\dot{{s}}(t)= -\left.\dfrac{d}{dt'}\left( \ln p_t(\bm{A}(t'))+\dfrac{1}{k_\text{B}T}\mathcal{H}(\bm{A}(t'))\right)\right|_{t'=t}.
\end{equation}
As discussed in Appendix~\ref{thermodynamics}, $k_\text{B}\left\langle\dot{{s}}(t) \right\rangle $ is nothing but the rate of change of total entropy of the system and the reservoir (see Eq.~\eqref{entpass}). According to Eq.~\eqref{rtau0}, the irreversible behavior of the system results from entropy production and from the asymmetric behavior of $p_t(\bm{A})$  under time reversal. 
If $p_t(\bm{s}\circ\bm{A})\neq p_t(\bm{A})$,
the system is instantaneously irreversible since
\begin{equation}\label{key}
\mathcal{R}_0(t)=\ln\dfrac{p_t(\bm{A}(t))}{p_t(\bm{s}\circ\bm{A}(t))}\neq0.
\end{equation}
Since $\mathcal{H}(\bm{s}\circ\bm{A})=\mathcal{H}(\bm{A})$, the equilibrium probability distribution $p_\text{eq}(\bm{A})\equiv\exp(-\mathcal{H}(\bm{A})/k_\text{B}T)/\mathcal{Z}$ always follows the symmetry property $p_\text{eq}(\bm{s}\circ\bm{A})= p_\text{eq}(\bm{A})$; it is a fundamental property of $p_\text{eq}(\bm{A})$. The nonzero value of $\left\langle \mathcal{R}_0(t)\right\rangle$ for an out-of-equilibrium system signifies that the system violates this symmetry. Note that $\left\langle \mathcal{R}_0(t)\right\rangle\ge0$. An example of such systems is as follows: consider a colloidal particle moving with a nonzero average velocity $\bm{v}_0$ and having the probability distribution $p_0(\bm{v})=C\exp(-(\bm{v}-\bm{v}_0)^2/2)$, at $t=0$. Under time reversal $\bm{v}\to -\bm{v}$, so \hsfn{$\bm{s}=\{-1,-1,-1\}$}. Hence $p_0(\bm{s}\circ\bm{v})\neq p_0(\bm{v})$.

For $p_t(\bm{s}\circ\bm{A})= p_t(\bm{A})$ case, $\mathcal{R}_0(t)=0$, so from Eq.~\eqref{rtau0},
\begin{equation}\label{dissfdot}
\dot{{s}}(t)=\lim_{\tau\to 0} \dfrac{\mathcal{R}_\tau(t)}{\tau}.
\end{equation} 
Thus, \hsn{the average rate} of change of $\mathcal{R}_\tau(t)$ with $\tau$ is \hsn{the same as} the rate of the total entropy production of the system and the reservoir; from Eq.~\eqref{thermsecc}, the second law of thermodynamics is evident, $\left\langle\dot{{s}}(t) \right\rangle >0$. \hsn{In the next subsection, we discuss a broad class of passive systems with $p_t(\bm{s}\circ\bm{A})= p_t(\bm{A})$.}

The form of the dissipation function used by Seifert \textit{et al.}~\cite{Seifert_prl_2005} is briefly discussed in Appendix~\ref{seif}.

\subsection{The dissipation function for quenched systems}~\label{qunchsys}
Here we consider that the system is initially in a thermodynamic equilibrium state and the state variables of the system $\bm{\beta}\equiv\{\beta_1,\beta_2,...,\beta_n\}$ are abruptly changed  at $t=0$. Then the system will start evolving towards the equilibrium state corresponding to the  modified values of $\bm{\beta}$. Writing the coarse-grained Hamiltonian of the system as the function of $\bm{\beta}$: $\mathcal{H}\equiv \mathcal{H}(\bm{A};\bm{\beta})$. Let $\bm{\beta}=\bm{\beta}_\text{I}$ at $t=0$ then 
\begin{equation}\label{key}
p_0(\bm{A})=\dfrac{1}{\mathcal{Z}(\bm{\beta}_\text{I})}\exp\left [-\dfrac{\mathcal{H}(\bm{A};\bm{\beta}_\text{I})}{k_\text{B}T}\right]
\end{equation}
From Eq.\eqref{dissf}, for a quench from  $\bm{\beta}=\bm{\beta}_\text{I}$ to $\bm{\beta}=\bm{\beta}_\text{F}$ at $t=0$, the dissipation function for the system takes the following form
\begin{eqnarray}\label{disstrns}
\nonumber
\mathcal{R}_\tau=\dfrac{1}{k_\text{B}T}\left[\mathcal{H}(\bm{A}(0);\bm{\beta}_\text{F})-\mathcal{H}(\bm{A}(0);\bm{\beta}_\text{I})\right.\\
   -\left.(\mathcal{H}(\bm{A}(\tau);\bm{\beta}_\text{F})-\mathcal{H}(\bm{A}(\tau);\bm{\beta}_\text{I})) \right].
\end{eqnarray}
We will now discuss an example of quenched systems.
\subsubsection{Colloidal particle in a harmonic potential well}
Consider a colloidal particle trapped in a harmonic potential $U=k \mathbf{r}^2/2$, where $k$ is the stiffness of the potential. Imagine that initially the particle is in thermodynamic equilibrium with $k=k_0$ and the value of $k$ is instantaneously changed from $k_0$ to $k_1$ at $t=0$~\cite{op_tweez_exp_prl_2004}.  Ignoring the kinetic energy, the coarse-grained Hamiltonian for this system would be simply $\mathcal{H}=U$.  Then, from Eq.~\eqref{disstrns}, the dissipation function for a trajectory between $\mathbf{r}=\mathbf{r}_0$ and $\mathbf{r}=\mathbf{r}_\tau$ in time $\tau$ is given by
\begin{equation}\label{trapdiss}
R_\tau=\dfrac{1}{2}(k_0-k_1)(\mathbf{r}_\tau^2-\mathbf{r}^2_0)
\end{equation}
The above expression was derived by Carberry \textit{et al.}~\cite{reid2004reversibility,op_tweez_exp_prl_2004} for spatially uniform diffusion constant. As we have considered the dependence of $\Gamma_{ij}$ on $\bm{A}$ in the derivation of $\mathcal{R}_\tau$, the above expression of the dissipation function is more general; it is valid for the systems having state dependent diffusion as well. In order to verify our prediction, we numerically solve the Langevin equation for a colloidal particle with the diffusion coefficient varying with position. For simplicity, we consider \hsn{the} 1D case. From Eq.~\eqref{gle2}, the overdamped Langevin equation for the colloidal particle reads
\begin{equation}\label{key}
\dfrac{dx}{dt}=-\dfrac{1}{k_\text{B}T}D(x)k x+(1-\alpha)\dfrac{dD(x)}{dx}
+\sqrt{2D(x)}\eta(t),
\end{equation}
with its discreate form (see Eq.~\eqref{geldisc})
\begin{eqnarray}\label{coldiscf}
\nonumber
x(t+dt)=x(t)-\dfrac{1}{k_\text{B}T}D(\bar{x}^\text{f})k \bar{x}^\text{f}&+&(1-\alpha)\left[ \dfrac{dD(x)}{dx}\right]_{\bar{x}^\text{f}}\\
&+&\sqrt{2D(\bar{x}^\text{f})dt}\eta_t,
\end{eqnarray}
where  $\bar{x}^\text{f}=\alpha x(t+dt)+(1-\alpha)x(t)$ and $D(x)$ is the state-dependent diffusion. \hsn{The above} equation is a self-consistent equation of $x(t+dt)$ for given $x(t)$.  There are many examples of the systems having state-dependent diffusion; e.g., a colloidal particle near a wall~\cite{lau2007state}. We here consider a hypothetical system having Gaussian profile of the diffusion coefficient:
\begin{equation}\label{key}
D(x)=D_0\exp\left( -\dfrac{x^2}{L^2}\right).
\end{equation}
\begin{figure}[h]
	\begin{center}
		\includegraphics[width=0.38\textwidth]{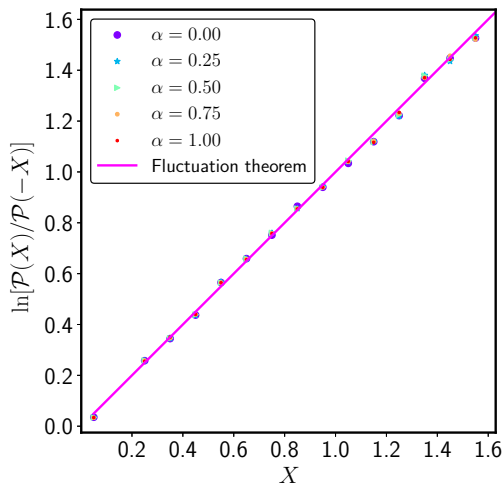}\\
		\caption{\hsn{$\ln[\mathcal{P}(X)/\mathcal{P}(-X)]$  vs $X$} for a 1D colloidal particle in  a potential well $U=k x^2/2$ with the diffusion coefficient $D(x)=D_0\exp (-x^2/L^2)$, where $\mathcal{P}$ is the probability distribution function for the dissipation function given by Eq.~\eqref{trapdiss}. The value of $k$ is suddenly changed from $k=k_0$ to $k=k_1$  at $t=0$. Here $L/\sqrt{k_1/k_\text{B}T}=1$, $k_1/k_0=4$, the time duration of the trajectory $\tau=3k_\text{B}T/D_0k_1$, and 8,00,000 no. of trajectories are used for the statistics.}
		\label{ftfig}
	\end{center}
\end{figure}
To obtain the trajectory of the particle, at each time step, we solve the  Eq.~\eqref{coldiscf} for $x(t+dt)$ with fixed point iteration method with the accuracy of $10^{-4}$.
In Fig.~\ref{ftfig}, we show \hsn{$\ln[\mathcal{P}(X)/\mathcal{P}(-X)]$  vs $X$}: clearly, the dissipation function given by Eq.~\eqref{trapdiss} obeys the fluctuation relation, for all the values of $\alpha$.

%
\section{The dissipation function for active systems}~\label{actdiss}
In this section, we consider the active systems~\cite{RevModPhys.85.1143} whose dynamics is governed by the  equations of motion having the following form:
\begin{equation}\label{key}
\dfrac{dA_i}{dt}=\mathcal{F}_i+\mathcal{X}_i+N_{ij}\xi_j(t),
\end{equation}
where the addition term $\mathcal{X}_i$ represents the active driving forces. \hsn{	
Due to the presence of the active forces, the active systems are always away from equilibrium. However, they can achieve a nonequilibrium steady state. 
Writing  ${\mathcal{X}}_i$ as  the sum of  two terms ${\mathcal{X}}^\text{s}_i$ and ${\mathcal{X}}^\text{a}_i$ such that $\bm{\mathcal{X}}^\text{s}(\bm{s}\circ\bm{A})= \bm{s}\circ \bm{\mathcal{X}}^\text{s}(\bm{A})$ and $\bm{\mathcal{X}}^\text{a}(\bm{s}\circ\bm{A})= -\bm{s}\circ \bm{\mathcal{X}}^\text{a}(\bm{A})$ (see Eqs.~\eqref{Xs} \&~\eqref{Xa}, Appendix~\ref{sdot}). It should be noted that $N_{ij}$ serves as a dummy matrix here as well because the form of $P$ will be the same as that in Eq.~\eqref{Pform}, except that $\mathcal{F}^\text{s}_i$ and $\mathcal{F}^\text{a}_i$ will have additional active components ${\mathcal{X}}^\text{s}_i$ and ${\mathcal{X}}^\text{a}_i$, respectively.}
Following the approach used in subsection~\ref{ratio}, we obtain the following expression of the dissipation function:
\begin{eqnarray}\label{rtauact}
\nonumber
\mathcal{R}_\tau(t)=\ln \dfrac{p_t(\bm{A}(t))}{p_t(\bm{s}\circ\bm{A}(t+\tau))}-\dfrac{1}{k_\text{B}T}\left[  \mathcal{H}(\bm{A}(t+\tau))\right. \\ \left. -\mathcal{H}(\bm{A}(t))\right]+\dfrac{1}{k_\text{B}T}\int^{t+\tau}_t w(t')dt',
\end{eqnarray}
where 
\begin{eqnarray}\label{stau}
\nonumber
w(t)&=& \dfrac{\partial \mathcal{H}(\bm{A}(t))}{\partial A_i}\mathcal{X}^\text{a}_i(\bm{A}(t))-k_\text{B}T\dfrac{\partial \mathcal{X}^\text{a}_i(\bm{A}(t))}{\partial A_i} \\&+&\hsn{({\Gamma^{\text{s}}}^{-1})_{ij}}(\bm{A}(t)) \mathcal{X}^\text{s}_j(\bm{A}(t))\left[\dfrac{dA_i}{dt}-\mathcal{Y}^\text{a}_i(\bm{A}(t))\right]
\end{eqnarray}
and
\begin{equation}\label{key}
\mathcal{Y}^\text{a}_i= \mathcal{X}^\text{a}_i +\dfrac{\partial \mathcal{H}}{\partial A_k}\Gamma^\text{a}_{ki}-k_\text{B}T\dfrac{\partial \Gamma^\text{a}_{ki}}{\partial A_k}.
\end{equation}
The integration in Eq.~\eqref{stau} is performed using midpoint rule. \hsn{In contrast to passive systems,  $\mathcal{R}_\tau(t)$ here depends on $\Gamma_{ij}$, though not on $\alpha$. Moreover, $\mathcal{R}_\tau(t)$ is trajectory-dependent,  so $\left\langle \mathcal{R}_\tau(t)\right\rangle $ is a function of $\alpha$ because the probability density $P$ of a trajectory depends on $\alpha$ (as in passive systems, see Eq.~\eqref{Pform}).} 
 The ensemble average of $w(t)$ is given by \hsn{(with an assumption, see Appendix~\ref{sdot})}
\begin{eqnarray}\label{key}
\nonumber
\left\langle w(t)\right\rangle &=&\left\langle \left( \dfrac{dA_i}{dt}-\dfrac{\partial \mathcal{H}}{\partial A_k}\Gamma^\text{a}_{ki}+k_\text{B}T\dfrac{\partial \Gamma^\text{a}_{ki}}{\partial A_k}\right) \hsn{({\Gamma^{\text{s}}}^{-1})_{ij}} \mathcal{X}^\text{s}_j\right\rangle\\&&-\left\langle \left( \dfrac{dA_i}{dt}- \mathcal{Y}^\text{a}_i\right)\hsn{({\Gamma^{\text{s}}}^{-1})_{ij}} \mathcal{X}^\text{a}_j \right\rangle,
\end{eqnarray}
where the first term is the average rate of work done by active force $\hsn{({\Gamma^{\text{s}}}^{-1})_{ij}} \mathcal{X}^\text{s}_j$ and the second term is by the active force $\hsn{({\Gamma^{\text{s}}}^{-1})_{ij}} \mathcal{X}^\text{a}_j$. So $w(t)$ can be interpreted as  the rate of work done by active forces along the trajectory at time $t$.
For $p_t(\bm{s}\circ\bm{A})= p_t(\bm{A})$, the rate of change of dissipation function with $\tau$ (see Eqs.~\eqref{dissfdot} and~\eqref{rtauact}) is given by
\begin{eqnarray}\label{disspumped}
\nonumber
\dot{{s}}(t)=\dfrac{1}{k_\text{B}T}w(t) -\left.\dfrac{d}{dt'}\left( \ln p_t(\bm{A}(t'))+\dfrac{1}{k_\text{B}T}\mathcal{H}(\bm{A}(t'))\right)\right|_{t'=t}.\\
\label{rdotdri}
\end{eqnarray}
and its average reads (see Appendix~\ref{sdot})
\begin{eqnarray}\label{sdotact}
\nonumber
\left\langle \dot{{s}}(t)\right\rangle=\dfrac{1}{k_\text{B}T}\left\langle \left(\dfrac{J_i(\bm{A},t)}{p_t(\bm{A})}-\mathcal{Y}^\text{a}_i(\bm{A}) \right)\hsn{({\Gamma^{\text{s}}}^{-1})_{ij}}(\bm{A})\right. \\
\times \left. \left( \dfrac{J_j(\bm{A},t)}{p_t(\bm{A})}-\mathcal{Y}^\text{a}_j(\bm{A})\right)\right\rangle,
\end{eqnarray}
where $J_i(\bm{A},t)$ is the probability current for $\bm{A}$~\cite{lau2007state}:
\begin{eqnarray}\label{key}
\nonumber
J_i(\bm{A},t)&=&\left( -{\Gamma^{\text{s}}}_{ij}(\bm{A})\dfrac{\partial \mathcal{H}(\bm{A})}{\partial A_j}+\mathcal{Y}^\text{a}_i(\bm{A}) +\mathcal{X}^\text{s}_j(\bm{A})\right) p_t(\bm{A})\\&&\qquad\qquad\qquad\qquad-k_\text{B}T\Gamma^{\text{s}}_{ij}(\bm{A})\dfrac{\partial p_t(\bm{A})}{dA_j}.
\end{eqnarray}	
 \hsn{Using Eq~\eqref{nn},} it is easy to show that $\left\langle \dot{{s}}(t)\right\rangle>0$,
as expected from the integrated fluctuation theorem~\eqref{ift}. Again,  $k_\text{B}\left\langle \dot{{s}}(t)\right\rangle$ is nothing but the rate of the total entropy production of the system and the reservoir (see Appendix~\ref{thermodynamics}).

As discussed for the passive systems in subsection~\ref{ftrsub}, for $p_t(\bm{s}\circ\bm{A})\neq p_t(\bm{A})$ case, 
time reversal asymmetry in $p_t(\bm{A})$  also contributes to irreversibility. \hsn{This contribution can also be observed in the stationary states of many active systems. 
Active systems with polar alignment~\cite{Bricard2013,hs_nk,PhysRevLett.110.208001,chatterjee2021inertia} are examples of this type of system; as the velocities of the particles are globally aligned, the velocity distribution is not an even function for these systems. The passive systems, being in equilibrium in their stationary states, cannot demonstrate this irreversibility.  } 

A few special cases for the active systems are as follows: (a)  if initially the system is in equilibrium \hsfn{(that is, $\mathcal{X}_i=0$)} and the active term $\mathcal{X}_i$ is switched on at $t=0$ then $p_0(\bm{A})=\exp(-\mathcal{H}(\bm{A})/k_\text{B}T)/\mathcal{Z}$. In this case, $\mathcal{R}_\tau$ is just the net work done by the active forces during the trajectory (see Eq.~\eqref{rtauact}):
\begin{equation}\label{key}
\mathcal{R}_\tau=\hsn{\dfrac{1}{k_\text{B}T}\int^{\tau}_0} w(t')dt'.
\end{equation}
(b) \hsn{In the stationary state (\textit{i.e.} \hsfn{in $t\to\infty$} limit), the time averaged work done by the active forces,  
\begin{equation}\label{key}
w_\text{av}= \lim_{\tau\to\infty}\dfrac{1}{\tau}\int^{t+\tau}_{t}dt'w(t'),
\end{equation}
is independent of $t$. So, in $\tau\to\infty$  limit, in  Eq.~\eqref{rtauact}, the last term  is proportional to  $\tau$ and we can ignore the first two terms. Then,
\begin{eqnarray}\label{key}
\mathcal{R}_\tau&\simeq&\dfrac{1}{k_\text{B}T}\int^{t+\tau}_t w(t')dt'.\\
&\simeq&\dfrac{1}{k_\text{B}T}w_\text{av}\tau.
\end{eqnarray}
From Eq.~\eqref{ft}, the probability distribution $\mathcal{P}_\text{w}(w_\text{av})$ of $w_\text{av}$ satisfies the relation
\begin{equation}\label{ssft}
X=\lim_{\tau\to\infty}k_\text{B}T\dfrac{1}{\tau}\ln \dfrac{\mathcal{P}_\text{w}(w_\text{av}=X)}{\mathcal{P}_\text{w}(w_\text{av}=-X)},
\end{equation}
This is called the steady-state fluctuation theorem~\cite{ft_evan_review}.}

\section{Conclusion}
Starting with the generic Langevin equations, using path integral approach, we first calculated the ratio of the probability densities of a trajectory and its time-reversed trajectory for passive systems using $\alpha$-discritization: it is independent of the value of $\alpha$. \hsn{Irrespective of the value of $\alpha$, the stationary solutions of  generic Langevin equations have time reversal symmetry, so the generic Langevin equations with any value of $\alpha$ describes a passive system.}
Next we calculated the dissipation function for the passive systems which is found to be independent of the trajectory of the system, it depends only on the intial and the final values of the dynamical variables of the system. Furthermore, it is not an explicit function  of coefficients of the generic Langevin equations.  We also verify the fluctuation theorem for a 1D particle trapped in a potential well whose stiffness is suddenly changed, with the state-dependent diffusion. Finally, we obtained the expression of the dissipation function for active systems and defined the work done by the active forces. For both passive and active systems, the average of the rate of change of dissipation function with the duration of the trajectory is just the entropy production rate of the system and the reservoir.

\appendix
\begin{widetext}
\section{The probability density of a trajectory for passive systems}\label{Appdf}
The generic Langevin equations for passive systems in discrete form (see Eq.~\eqref{geldisc}):
\begin{equation}\label{Append1}
dA_i(l)=\epsilon\mathcal{F}_i(\bar{\bm{A}}^\text{f}_l)+\sqrt{\epsilon}N_{ij}(\bar{\bm{A}}^\text{f}_l)\xi_j^{l},
\end{equation}
\hsn{where} $dA_i(l)\equiv A_i(\epsilon l)-A_i(\epsilon (l-1))$, $\bar{\bm{A}}^\text{f}_l\equiv \alpha \bm{A}(\epsilon l)+(1-\alpha) \bm{A}(\epsilon (l-1))$, \hsn{and
\begin{equation}\label{F}
\mathcal{F}_i\equiv-\Gamma_{ij}\dfrac{\partial \mathcal{H}}{\partial A_j}+k_\text{B}T\dfrac{\partial \Gamma_{ij}}{\partial A_j}-\alpha N_{lj}  \dfrac{\partial N_{ij}}{\partial A_l}.
\end{equation}	
Solving} the above equations for $\xi_i^{l}$, we obtain
\begin{equation}\label{Appendxi}
\xi_i^{l}=\dfrac{1}{\sqrt{\epsilon}}\hsn{(N^{-1})_{ij}}(\bar{\bm{A}}^\text{f}_l) (dA_j(l)-\epsilon\mathcal{F}_j(\bar{\bm{A}}^\text{f}_l)).
\end{equation}
Since  $\xi_i^{l}$  are the uncorrelated series of random numbers having normal distribution with zero mean and variance one, the probability density function of a trajectory of the system $(\bm{A}_0,\bm{A}_1,\bm{A}_2,.....\bm{A}_N)$ (here $\bm{A}_l\equiv\bm{A}(l\epsilon)$) between $t=0$ and $t=\tau \equiv N \epsilon$ is given by~\cite{soni2019flocks}
\begin{equation}\label{AppendP1}
P=p_0(\bm{A}_0)\left|  \mathcal{J}\right| \prod^{N}_{l=1} \dfrac{1}{(2\pi )^{n/2}}\exp\left[-\dfrac{1}{2\epsilon} \left[dA_i(l)- \epsilon\mathcal{F}_i(\bar{\bm{A}}^\text{f}_l)\right] \hsn{(N^{-1})_{ki}}(\bar{\bm{A}}^\text{f}_l)\hsn{(N^{-1})_{kj}}(\bar{\bm{A}}^\text{f}_l)\left[dA_j(l)- \epsilon\mathcal{F}_j(\bar{\bm{A}}^\text{f}_l)\right]\right],
\end{equation} 
where $p_0(\bm{A})$ is the probability distribution of $\bm{A}$ at $t=0$ and $\mathcal{J}$ is the Jacobean determinant for the transformation of the variables of  the probability density function from the $\xi_i^{l}$ to $A_j(\epsilon m)$. From  Eq.~\eqref{Appendxi}, the $Nn\times Nn$ Jacobean matrix for the variable transformation is given by
\begin{eqnarray}\label{key}
\nonumber
\mathcal{J}^{il}_{jm}&=&\dfrac{\partial \xi_i^{l}}{\partial A_j(\epsilon m)}\\
\nonumber
&=&\dfrac{1}{\epsilon^{1/2}}\left[ \left[ \left[ \dfrac{\partial \hsn{(N^{-1})_{ik}}}{\partial A_j}\right]_{\bar{\bm{A}}^\text{f}_l}dA_k(l)-\epsilon\left[ \dfrac{\partial \hsn{(N^{-1})_{ik}}}{\partial A_j}\mathcal{F}_k+ \dfrac{\partial \mathcal{F}_k}{\partial A_j} \hsn{(N^{-1})_{ik}}\right]_{\bar{\bm{A}}^\text{f}_l}\right] (\alpha\delta_{lm}+(1-\alpha)\delta_{(l-1)m})\right.\\
\nonumber
&&\Bigg.+ \hsn{(N^{-1})_{ij}}(\bar{\bm{A}}^\text{f}_l)( \delta_{lm}-\delta_{(l-1)m})\Bigg] \\
\nonumber
&=&\dfrac{1}{\epsilon^{1/2}}\left[ \left[ \left[ \dfrac{\partial \hsn{(N^{-1})_{ik}}}{\partial A_j}\right]_{\bar{\bm{A}}^\text{f}_l}\left( dA_k(l)-\epsilon\mathcal{F}_k(\bar{\bm{A}}^\text{f}_l)\right) -\epsilon\left[  \dfrac{\partial \mathcal{F}_k}{\partial A_j} \hsn{(N^{-1})_{ik}}\right]_{\bar{\bm{A}}^\text{f}_l}\right] (\alpha\delta_{lm}+(1-\alpha)\delta_{(l-1)m})\right. \\
&&\Bigg.+ \hsn{(N^{-1})_{ij}}(\bar{\bm{A}}^\text{f}_l)( \delta_{lm}-\delta_{(l-1)m})\Bigg].
\end{eqnarray}
The above matrix is a block triangular matrix of $n\times n$ submatrices with fixed ($l,m$), so its determinant will be the multiplication of all the diagonal submatrices (i.e. with $l=m$):
\begin{equation}\label{key}
\mathcal{J}=\prod^{N}_{l=1}\dfrac{1}{\epsilon^{n/2}} \hsfn{\text{det}\left(\bm{M}(l)\right) }  ,
\end{equation}
where
\begin{eqnarray}\label{key}
\nonumber
M_{ij}(l)&=& \hsn{(N^{-1})_{ij}}(\bar{\bm{A}}^\text{f}_l)+\alpha  \left[ \dfrac{\partial \hsn{(N^{-1})_{ik}}}{\partial A_j}\right]_{\bar{\bm{A}}^\text{f}_l}\left( dA_k(l)-\epsilon\mathcal{F}_k(\bar{\bm{A}}^\text{f}_l)\right) -\epsilon\alpha\left[  \dfrac{\partial \mathcal{F}_k}{\partial A_j} \hsn{(N^{-1})_{ik}}\right]_{\bar{\bm{A}}^\text{f}_l} \\ 
\label{a5}
&=&\hsn{(N^{-1})_{ip}}(\bar{\bm{A}}^\text{f}_l)\left[  \delta_{pj}+\alpha \left[ N_{pq} \dfrac{\partial \hsn{(N^{-1})_{qk}}}{\partial A_j}\right]_{\bar{\bm{A}}^\text{f}_l}\left( dA_k(l)-\epsilon\mathcal{F}_k(\bar{\bm{A}}^\text{f}_l)\right)-\alpha\epsilon\left[\dfrac{\partial \mathcal{F}_p}{\partial A_j}\right]_{\bar{\bm{A}}^\text{f}_l} \right]. 
\end{eqnarray}
\hsn{Using the power series expression of $ \ln ( \hsfn{\text{det}\left(\bm{I}+\delta \bm{\mathcal{B}}\right) } )$  for any matrix $\bm{\mathcal{B}}$ in $\delta\to0$ limit (such that $||\delta\bm{\mathcal{B}}||<1$), that is,
\begin{equation} \label{logprop}
\ln( \hsfn{\text{det}\left(\bm{I}+\delta \bm{\mathcal{B}}\right) }) =\left[ \text{Tr}[\bm{\mathcal{B}}]\delta-\dfrac{1}{2}\text{Tr}[\bm{\mathcal{B}}\cdot\bm{\mathcal{B}}]\delta^2+\mathcal{O}(\delta^{3})\right],
\end{equation}
the determinant of $\bm{M}(l)$ can be written as}
\begin{eqnarray}\label{key}
\nonumber
 \hsfn{\text{det}\left(\bm{M}(l)\right) }
&=& \hsfn{\text{det}\left(\bm{N}^{-1}(\bar{\bm{A}}^\text{f}_l)\right) } \exp \left[\alpha\left[ \left[ N_{jq} \dfrac{\partial \hsn{(N^{-1})_{qk}}}{\partial A_j}\right]_{\bar{\bm{A}}^\text{f}_l}\left( dA_k(l)-\epsilon\mathcal{F}_k(\bar{\bm{A}}^\text{f}_l)\right)-\epsilon\left[\dfrac{\partial \mathcal{F}_j}{\partial A_j}\right]_{\bar{\bm{A}}^\text{f}_l}\right]\right.\\
\nonumber
&& \left. \hsn{-\dfrac{1}{2}\alpha^2\left[ N_{pq}\dfrac{\partial (N^{-1})_{qk}}{\partial A_j}N_{jr}\dfrac{\partial (N^{-1})_{ri}}{\partial A_p}\right]_{\bar{\bm{A}}^\text{f}_l} dA_k(l)dA_i(l)+\mathcal{O}(\epsilon^{3/2})}\right],\\
\nonumber
&=&\hsfn{\text{det}\left(\bm{N}^{-1}(\bar{\bm{A}}^\text{f}_l)\right) } \exp \left[-\alpha\left[ \left[ \hsn{(N^{-1})_{qk}} \dfrac{\partial N_{jq} }{\partial A_j}\right]_{\bar{\bm{A}}^\text{f}_l}\left( dA_k(l)-\epsilon\mathcal{F}_k(\bar{\bm{A}}^\text{f}_l)\right)+\epsilon\left[\dfrac{\partial \mathcal{F}_j}{\partial A_j}\right]_{\bar{\bm{A}}^\text{f}_l}\right]\right.\\
&& \left. \hsn{-\dfrac{1}{2}\alpha^2\epsilon\left[ \dfrac{\partial N_{pm} }{\partial A_j}\dfrac{\partial N_{jm}}{\partial A_p}\right]_{\bar{\bm{A}}^\text{f}_l} +\mathcal{O}(\epsilon^{3/2})}\right].
\end{eqnarray}
\hsn{Note that $dA_k(l)$ has a $\epsilon^{1/2}$-term, so $dA_i(l)dA_k(l)$ is of the order of $\epsilon$. Relations~\eqref{D1} \&~\eqref{D66}  (see Appendix~\ref{indent}) have been used to get the last term of the above equation.}
Eq.~\eqref{a5} then reads
\begin{eqnarray}\label{key}
\nonumber
\mathcal{J}&=&\prod^{N}_{l=1}\dfrac{1}{\epsilon^{n/2}}\hsfn{\text{det}\left(\bm{N}^{-1}(\bar{\bm{A}}^\text{f}_l)\right) }\exp \left[-\alpha \left[ \left[  \hsn{(N^{-1})_{qk}}\dfrac{\partial N_{jq}}{\partial A_j}\right]_{\bar{\bm{A}}^\text{f}_l}\left( dA_k(l)-\epsilon\mathcal{F}_k(\bar{\bm{A}}^\text{f}_l)\right)+\epsilon\left[\dfrac{\partial \mathcal{F}_j}{\partial A_j}\right]_{\bar{\bm{A}}^\text{f}_l}\right]\right.\\
&& \left. \hsn{-\dfrac{1}{2}\alpha^2\epsilon\left[ \dfrac{\partial N_{pm} }{\partial A_j}\dfrac{\partial N_{jm}}{\partial A_p}\right]_{\bar{\bm{A}}^\text{f}_l} +\mathcal{O}(\epsilon^{3/2})}\right].
\end{eqnarray}
Substituting the above expression of $\mathcal{J}$ into Eq.\eqref{AppendP1}:
\begin{eqnarray}\label{AppendP21}
\nonumber
P&=&p_0(\bm{A}_0)\prod^{N}_{l=1}\left\lbrace   \dfrac{1}{(2\pi \epsilon)^{n/2}}\hsn{\left|\text{det}\left(   \bm{N}^{-1}(\bar{\bm{A}}^\text{f}_l)\right) \right|}\exp\left[-\dfrac{1}{2\epsilon} \left[dA_i(l)- \epsilon\mathcal{F}_i(\bar{\bm{A}}^\text{f}_l)\right] \hsn{(N^{-1})_{ki}}(\bar{\bm{A}}^\text{f}_l)\hsn{(N^{-1})_{kj}}(\bar{\bm{A}}^\text{f}_l)\left[dA_j(l)- \epsilon\mathcal{F}_j(\bar{\bm{A}}^\text{f}_l)\right]\right]\right. \\
&&\left. \times  \exp \left[-\alpha \left[ \left( dA_i(l)-\epsilon\mathcal{F}_i(\bar{\bm{A}}^\text{f}_l)\right)\left[  \hsn{(N^{-1})_{ji}}\dfrac{\partial N_{kj}}{\partial A_k}\right]_{\bar{\bm{A}}^\text{f}_l}+\epsilon\left[\dfrac{\partial \mathcal{F}_i}{\partial A_i}\right]_{\bar{\bm{A}}^\text{f}_l}\right] \hsn{-\dfrac{1}{2}\alpha^2\epsilon\left[ \dfrac{\partial N_{pm} }{\partial A_j}\dfrac{\partial N_{jm}}{\partial A_p}\right]_{\bar{\bm{A}}^\text{f}_l} +\mathcal{O}(\epsilon^{3/2})}\right]\right\rbrace.
\end{eqnarray} 
From Eq.~\eqref{gamma}, $\hsn{(N^{-1})_{ki}}\hsn{(N^{-1})_{kj}}=\hsn{(\Gamma^{s}}^{-1})_{ij}/2 k_\text{B}T$, so
\begin{eqnarray}\label{AppendP22}
\nonumber
P&=&p_0(\bm{A}_0)\prod^{N}_{l=1}\left\lbrace   \dfrac{\hsn{(2k_\text{B}T)^{-1/2}}}{(2\pi \epsilon)^{n/2}}\hsn{\text{det}\left(  \bm{\Gamma^\text{s}}(\bar{\bm{A}}^\text{f}_l)\right)^{-1/2}} \exp\left[-\dfrac{1}{4\epsilon k_\text{B}T} \left[dA_i(l)- \epsilon\mathcal{F}_i(\bar{\bm{A}}^\text{f}_l)\right] {\hsn{(\Gamma^{s}}^{-1})_{ij}}(\bar{\bm{A}}^\text{f}_l)\left[dA_j(l)- \epsilon\mathcal{F}_j(\bar{\bm{A}}^\text{f}_l)\right]\right]\right. \\
\nonumber
&&\left. \times  \exp \left[-\alpha \left[ \left( dA_i(l)-\epsilon\mathcal{F}_i(\bar{\bm{A}}^\text{f}_l)\right)\left[  \hsn{(N^{-1})_{ji}}\dfrac{\partial N_{kj}}{\partial A_k}\right]_{\bar{\bm{A}}^\text{f}_l}+\epsilon\left[\dfrac{\partial \mathcal{F}_i}{\partial A_i}\right]_{\bar{\bm{A}}^\text{f}_l}\right] \hsn{-\dfrac{1}{2}\alpha^2\epsilon\left[ \dfrac{\partial N_{pm} }{\partial A_j}\dfrac{\partial N_{jm}}{\partial A_p}\right]_{\bar{\bm{A}}^\text{f}_l} +\mathcal{O}(\epsilon^{3/2})}\right]\right\rbrace .\\
\end{eqnarray} 
\hsn{Now let us break $\mathcal{F}_i$ into two terms,
\begin{equation}\label{F}
\mathcal{F}^0_i=-\Gamma_{ij}\dfrac{\partial \mathcal{H}}{\partial A_j}+k_\text{B}T\dfrac{\partial \Gamma_{ij}}{\partial A_j}
\end{equation}
and 
\begin{equation}\label{key}
 \mathcal{F}^\text{N}_i=-\alpha N_{lj}  \dfrac{\partial N_{ij}}{\partial A_l}.
\end{equation}
\hsfn{
Replacing $\mathcal{F}_i$ by $\mathcal{F}^0_i+\mathcal{F}^\text{N}_i$ in Eq.~\eqref{AppendP22}:
\begin{eqnarray}\label{AppendP22n}
\nonumber
P&=&p_0(\bm{A}_0)\prod^{N}_{l=1}\left\lbrace   \dfrac{(2k_\text{B}T)^{-1/2}}{(2\pi \epsilon)^{n/2}}\text{det}\left(  \bm{\Gamma^\text{s}}(\bar{\bm{A}}^\text{f}_l)\right)^{-1/2} \exp\left[-\dfrac{1}{4\epsilon k_\text{B}T} \left[dA_i(l)- \epsilon\mathcal{F}^0_i(\bar{\bm{A}}^\text{f}_l)\right] {{(\Gamma^{s}}^{-1})_{ij}}(\bar{\bm{A}}^\text{f}_l)\left[dA_j(l)- \epsilon\mathcal{F}^0_j(\bar{\bm{A}}^\text{f}_l)\right]\right]\right. \\
\nonumber
&&\times  \exp \left[-\alpha \left[ \left( dA_i(l)-\epsilon\mathcal{F}^0_i(\bar{\bm{A}}^\text{f}_l)\right)\left[  (N^{-1})_{ji}\dfrac{\partial N_{kj}}{\partial A_k}\right]_{\bar{\bm{A}}^\text{f}_l}+\epsilon\left[\dfrac{\partial \mathcal{F}^0_i}{\partial A_i}\right]_{\bar{\bm{A}}^\text{f}_l}\right] \right]\\
\nonumber
&&\times\exp\left[\dfrac{1}{2 k_\text{B}T} \mathcal{F}^\text{N}_i(\bar{\bm{A}}^\text{f}_l) {{(\Gamma^{s}}^{-1})_{ij}}(\bar{\bm{A}}^\text{f}_l)\left[dA_j(l)- \epsilon\mathcal{F}^0_j(\bar{\bm{A}}^\text{f}_l)\right]-\dfrac{\epsilon}{4 k_\text{B}T}\left[ \mathcal{F}^\text{N}_i {{(\Gamma^{s}}^{-1})_{ij}}\mathcal{F}^\text{N}_j\right]_{\bar{\bm{A}}^\text{f}_l}\right]\\
&&\left. \times  \exp \left[\alpha\epsilon \left[ \mathcal{F}^\text{N}_i (N^{-1})_{ji}\dfrac{\partial N_{kj}}{\partial A_k}-\dfrac{\partial \mathcal{F}^\text{N}_i}{\partial A_i}\right]_{\bar{\bm{A}}^\text{f}_l} -\dfrac{1}{2}\alpha^2\epsilon\left[ \dfrac{\partial N_{pm} }{\partial A_j}\dfrac{\partial N_{jm}}{\partial A_p}\right]_{\bar{\bm{A}}^\text{f}_l} +\mathcal{O}(\epsilon^{3/2})\right]\right\rbrace .
\end{eqnarray}
Using Eq.~\eqref{nn} (that is, $N_{ik}N_{jk}=2 k_\text{B}T \Gamma^\text{s}_{ij}$), one can write   
\begin{eqnarray}\label{key}
\nonumber
(N^{-1})_{ji}\dfrac{\partial N_{kj}}{\partial A_k}&=&\delta_{im}(N^{-1})_{jm}\dfrac{\partial N_{kj}}{\partial A_k}\\
\nonumber
&=&({\Gamma^{s}}^{-1})_{ip}{\Gamma^{s}}_{pm}(N^{-1})_{jm}\dfrac{\partial N_{kj}}{\partial A_k}\\
\nonumber
&=&\dfrac{1}{2k_\text{B}T}({\Gamma^{s}}^{-1})_{ip}N_{pj}\dfrac{\partial N_{kj}}{\partial A_k}\\
\nonumber
&=&\dfrac{1}{2k_\text{B}T}({\Gamma^{s}}^{-1})_{ip}\left( \dfrac{\partial (N_{kj}N_{pj})}{\partial A_k}- N_{kj}\dfrac{\partial N_{pj}}{\partial A_k}\right) \\
&=&({\Gamma^{s}}^{-1})_{ip}\dfrac{\partial {\Gamma^{s}}_{pk}}{\partial A_k}+ \dfrac{1}{2k_\text{B}T\alpha}({\Gamma^{s}}^{-1})_{ip} \mathcal{F}^\text{N}_p.
\end{eqnarray}
With the above expression, Eq.~\eqref{AppendP22n} reduces to
 }
\begin{eqnarray}\label{AppendP221}
\nonumber
P&=&p_0(\bm{A}_0)\prod^{N}_{l=1}\left\lbrace   \dfrac{(2k_\text{B}T)^{-1/2}}{(2\pi \epsilon)^{n/2}}\text{det}\left(  \bm{\Gamma^\text{s}}(\bar{\bm{A}}^\text{f}_l)\right)^{-1/2} \exp\Biggl[-\dfrac{1}{4\epsilon k_\text{B}T} \left[dA_i(l)- \epsilon\mathcal{F}^0_i(\bar{\bm{A}}^\text{f}_l)\right] {\hsn{(\Gamma^{s}}^{-1})_{ij}}(\bar{\bm{A}}^\text{f}_l)\left[dA_j(l)- \epsilon\mathcal{F}^0_j(\bar{\bm{A}}^\text{f}_l)\right]\Biggr. \right. \\
\nonumber
&&\left. -\alpha \left[ \left( dA_i(l)-\epsilon\mathcal{F}^0_i(\bar{\bm{A}}^\text{f}_l)\right)\left[  ({\Gamma^{s}}^{-1})_{ij}\dfrac{\partial \Gamma^{s}_{jk}}{\partial A_k}\right]_{\bar{\bm{A}}^\text{f}_l}+\epsilon\left[\dfrac{\partial \mathcal{F}^0_i}{\partial A_i}\right]_{\bar{\bm{A}}^\text{f}_l}\right]\right] \\
\nonumber
&&\times\left.   \exp \left[\dfrac{\epsilon}{4 k_\text{B}T}\left[\mathcal{F}^\text{N}_i({\Gamma^{s}}^{-1})_{ij}\left( \mathcal{F}^\text{N}_j+4 \alpha k_\text{B}T \dfrac{\partial \Gamma^{s}_{jk}}{\partial A_k}\right)  -4\alpha k_\text{B}T\dfrac{\partial \mathcal{F}^\text{N}_i}{\partial A_i}-2 k_\text{B}T\alpha^2  \dfrac{\partial N_{pm} }{\partial A_j}\dfrac{\partial N_{jm}}{\partial A_p}\right]_{\bar{\bm{A}}^\text{f}_l} +\mathcal{O}(\epsilon^{3/2})\right]\right\rbrace .\\
\end{eqnarray} 
\hsn{Using the relation $N_{ik}N_{jk}=2 k_\text{B}T \Gamma^\text{s}_{ij}$, further simplifying the last term of the above equation yields 
\begin{eqnarray}\label{AppendP221n}
\nonumber
P&=&p_0(\bm{A}_0)\prod^{N}_{l=1}\left\lbrace   \dfrac{(2k_\text{B}T)^{-1/2}}{(2\pi \epsilon)^{n/2}}\text{det}\left(  \bm{\Gamma^\text{s}}(\bar{\bm{A}}^\text{f}_l)\right)^{-1/2} \exp\Biggl[-\dfrac{1}{4\epsilon k_\text{B}T} \left[dA_i(l)- \epsilon\mathcal{F}^0_i(\bar{\bm{A}}^\text{f}_l)\right] {\hsn{(\Gamma^{s}}^{-1})_{ij}}(\bar{\bm{A}}^\text{f}_l)\left[dA_j(l)- \epsilon\mathcal{F}^0_j(\bar{\bm{A}}^\text{f}_l)\right]\Biggr. \right. \\
\nonumber
&&\left. -\alpha \left[ \left( dA_i(l)-\epsilon\mathcal{F}^0_i(\bar{\bm{A}}^\text{f}_l)\right)\left[  ({\Gamma^{s}}^{-1})_{ij}\dfrac{\partial \Gamma^{s}_{jk}}{\partial A_k}\right]_{\bar{\bm{A}}^\text{f}_l}+\epsilon\left[\dfrac{\partial \mathcal{F}^0_i}{\partial A_i}\right]_{\bar{\bm{A}}^\text{f}_l}\right]\right] \\
&&\times\left.   \exp \left[\alpha^2\epsilon k_\text{B}T\left[\dfrac{\partial^2 \Gamma^{s}_{ij}}{\partial A_i\partial A_j} -\dfrac{\partial \Gamma^{s}_{ik}}{\partial A_k}({\Gamma^{s}}^{-1})_{ij} \dfrac{\partial \Gamma^{s}_{jp}}{\partial A_p}\right]_{\bar{\bm{A}}^\text{f}_l} +\mathcal{O}(\epsilon^{3/2})\right]\right\rbrace.
\end{eqnarray}		}
We further split $\mathcal{F}^0_i$ into the two  terms,
\begin{equation}\label{Fs1}
\mathcal{F}^\text{s}_i(\bm{A})=-\Gamma^\text{s}_{ij}\dfrac{\partial \mathcal{H}}{\partial A_j}+k_\text{B}T\dfrac{\partial \Gamma^\text{s}_{ij}}{\partial A_j}
\end{equation}
and
\begin{equation}\label{Fs2}
\mathcal{F}^a_i(\bm{A})=-\Gamma^a_{ij}\dfrac{\partial \mathcal{H}}{\partial A_j}+k_\text{B}T\dfrac{\partial \Gamma^\text{a}_{ij}}{\partial A_j},
\end{equation}
such that, under time reversal (see ~\eqref{Fprop1} and ~\eqref{Fprop2}),
\begin{eqnarray}
\label{Fprop1s}
&&\bm{\mathcal{F}}^\text{s}(\bm{A})\to\bm{\mathcal{F}}^\text{s}(\bm{s}\circ\bm{A})=\bm{s}\circ \bm{\mathcal{F}}^\text{s}(\bm{A}),\\
\label{Fprop2s}
&&\bm{\mathcal{F}}^\text{a}(\bm{A})\to\bm{\mathcal{F}}^\text{a}(\bm{s}\circ\bm{A})=-\bm{s}\circ \bm{\mathcal{F}}^\text{a}(\bm{A}).
\end{eqnarray}
Eq.~\eqref{AppendP221} then becomes  
\begin{eqnarray}\label{AppendP222}
\nonumber
P&=&p_0(\bm{A}_0)\prod^{N}_{l=1}\left\lbrace    \dfrac{(2k_\text{B}T)^{-1/2}}{(2\pi \epsilon)^{n/2}}\exp\Biggl[-\dfrac{1}{4\epsilon k_\text{B}T} \left[dA_i(l)-\epsilon\mathcal{F}^\text{s}_i(\bar{\bm{A}}^\text{f}_l)-\epsilon\mathcal{F}^\text{a}_i(\bar{\bm{A}}^\text{f}_l)\right] {\hsn{(\Gamma^{s}}^{-1})_{ij}}(\bar{\bm{A}}^\text{f}_l)\left[dA_j(l)-\epsilon\mathcal{F}^\text{s}_j(\bar{\bm{A}}^\text{f}_l)-\epsilon\mathcal{F}^\text{a}_j(\bar{\bm{A}}^\text{f}_l)\right]\Biggr. \right. \\
\nonumber
&&\left. -\alpha \left[ \left( dA_i(l)-\epsilon\mathcal{F}^\text{s}_i(\bar{\bm{A}}^\text{f}_l)-\epsilon\mathcal{F}^\text{a}_i(\bar{\bm{A}}^\text{f}_l)\right)\left[  ({\Gamma^{s}}^{-1})_{ij}\dfrac{\partial \Gamma^{s}_{jk}}{\partial A_k}\right]_{\bar{\bm{A}}^\text{f}_l}+\epsilon\left[\dfrac{\partial \mathcal{F}^\text{s}_i}{\partial A_i}+\dfrac{\partial \mathcal{F}^\text{a}_i}{\partial A_i}\right]_{\bar{\bm{A}}^\text{f}_l}\right]\right]
 \text{det}\left(  \bm{\Gamma^\text{s}}(\bar{\bm{A}}^\text{f}_l)\right)^{-1/2} \\
&&\times\left.   \exp \left[\alpha^2\epsilon k_\text{B}T\left[\dfrac{\partial^2 \Gamma^{s}_{ij}}{\partial A_i\partial A_j} -\dfrac{\partial \Gamma^{s}_{ik}}{\partial A_k}({\Gamma^{s}}^{-1})_{ij} \dfrac{\partial \Gamma^{s}_{jp}}{\partial A_p}\right]_{\bar{\bm{A}}^\text{f}_l} +\mathcal{O}(\epsilon^{3/2})\right]\right\rbrace.
\end{eqnarray}
Clearly, for given $\Gamma^\text{s}_{ij}$, $P$ is independent of the choice of $N_{ij}$.
} 
 \section{The probability density for the time-reversed trajectory}\label{revp} 
 As the time-reversed  trajectory of the trajectory  $(\bm{A}_0,\bm{A}_1,\bm{A}_2,.....\bm{A}_N)$  is $(\bm{s}\circ\bm{A}_N,\bm{s}\circ\bm{A}_{N-1},.....\bm{s}\circ\bm{A}_1)$,  under time reversal, $\bm{A}_l \to \bm{s}\circ\bm{A}_{N-l}$ and therefore,
\begin{eqnarray}\label{key}
\nonumber
dA_i(l)&=& A_i(\epsilon l)-A_i(\epsilon (l-1))\\\nonumber&\to& s_i(A_i(\epsilon (N-l))-A_i(\epsilon (N-l+1))\\  
&\to&-s_idA_i(N-l+1),
\end{eqnarray} 
\hsn{(Einstein’s convention is not used here)} and
\begin{eqnarray}\label{key}
\nonumber
\bar{\bm{A}}^\text{f}_l&=&\alpha \bm{A}_l+(1-\alpha) \bm{A}_{l-1}\\\nonumber&\to&(\alpha \bm{s}\circ\bm{A}_{N-l}+(1-\alpha) \bm{s}\circ\bm{A}_{N-l+1})\\
&\to&\bm{s}\circ\bar{\bm{A}}^\text{r}_{N-l+1},
\end{eqnarray}
where $\bar{\bm{A}}^\text{r}_l\equiv(1-\alpha) \bm{A}(l)+\alpha \bm{A}(l-1)$.
With the above transformations, using the relation  \hsn{$\Gamma^\text{s}_{ij}=s_is_j\Gamma^\text{s}_{ij}$ and Eqs.~\eqref{Fprop1s}, ~\eqref{Fprop2s} \&~\eqref{AppendP222}, we obtain the following expression of the probability density of the time-reversed trajectory:
\begin{eqnarray}\label{AppendP223}
\nonumber
P_r&=&\prod^{N}_{l=1}\left\lbrace  \dfrac{(2k_\text{B}T)^{-1/2}}{(2\pi \epsilon)^{n/2}}\exp\Biggl[-\dfrac{1}{4\epsilon k_\text{B}T} \left[-dA_i(l')-\epsilon\mathcal{F}^\text{s}_i(\bar{\bm{A}}^\text{r}_{l'})+\epsilon\mathcal{F}^\text{a}_i(\bar{\bm{A}}^\text{r}_{l'})\right] {\hsn{(\Gamma^{s}}^{-1})_{ij}}(\bar{\bm{A}}^\text{r}_{l'})\left[-dA_j(l')-\epsilon\mathcal{F}^\text{s}_j(\bar{\bm{A}}^\text{r}_{l'})+\epsilon\mathcal{F}^\text{a}_j(\bar{\bm{A}}^\text{r}_{l'})\right]\Biggr. \right. \\
\nonumber
&&\left. -\alpha \left[ \left( -dA_i(l')-\epsilon\mathcal{F}^\text{s}_i(\bar{\bm{A}}^\text{r}_{l'})+\epsilon\mathcal{F}^\text{a}_i(\bar{\bm{A}}^\text{r}_{l'})\right)\left[  ({\Gamma^{s}}^{-1})_{ij}\dfrac{\partial \Gamma^{s}_{jk}}{\partial A_k}\right]_{\bar{\bm{A}}^\text{r}_{l'}}+\epsilon\left[\dfrac{\partial \mathcal{F}^\text{s}_i}{\partial A_i}-\dfrac{\partial \mathcal{F}^\text{a}_i}{\partial A_i}\right]_{\bar{\bm{A}}^\text{r}_{l'}}\right]\right]
 \text{det}\left(  \bm{\Gamma^\text{s}}(\bar{\bm{A}}^\text{r}_{l'})\right)^{-1/2} \\
&&\times\left.   \exp \left[\alpha^2\epsilon k_\text{B}T\left[\dfrac{\partial^2 \Gamma^{s}_{ij}}{\partial A_i\partial A_j} -\dfrac{\partial \Gamma^{s}_{ik}}{\partial A_k}({\Gamma^{s}}^{-1})_{ij} \dfrac{\partial \Gamma^{s}_{jp}}{\partial A_p}\right]_{\bar{\bm{A}}^\text{r}_{l'}} +\mathcal{O}(\epsilon^{3/2})\right]\right\rbrace p_0(\bm{s}\circ\bm{A}_N),
\end{eqnarray}}
where $l'=N-l+1$.
In the above equation, the index $l'$ runs from $N$ to 1 so we can replace $\prod^N_{l=1}$ by $\prod^1_{l'=N}\equiv\prod^N_{l'=1}$. Hence 
 \hsn{\begin{eqnarray}\label{AppendP223}
\nonumber
P_r&=&\prod^{N}_{l=1}\left\lbrace   \dfrac{(2k_\text{B}T)^{-1/2}}{(2\pi \epsilon)^{n/2}}\exp\Biggl[-\dfrac{1}{4\epsilon k_\text{B}T} \left[-dA_i(l)-\epsilon\mathcal{F}^\text{s}_i(\bar{\bm{A}}^\text{r}_{l})+\epsilon\mathcal{F}^\text{a}_i(\bar{\bm{A}}^\text{r}_{l})\right] {\hsn{(\Gamma^{s}}^{-1})_{ij}}(\bar{\bm{A}}^\text{r}_{l})\left[-dA_j(l)-\epsilon\mathcal{F}^\text{s}_j(\bar{\bm{A}}^\text{r}_{l})+\epsilon\mathcal{F}^\text{a}_j(\bar{\bm{A}}^\text{r}_{l})\right]\Biggr. \right. \\
\nonumber
&&\left. -\alpha \left[ \left( -dA_i(l)-\epsilon\mathcal{F}^\text{s}_i(\bar{\bm{A}}^\text{r}_{l})+\epsilon\mathcal{F}^\text{a}_i(\bar{\bm{A}}^\text{r}_{l})\right)\left[  ({\Gamma^{s}}^{-1})_{ij}\dfrac{\partial \Gamma^{s}_{jk}}{\partial A_k}\right]_{\bar{\bm{A}}^\text{r}_{l}}+\epsilon\left[\dfrac{\partial \mathcal{F}^\text{s}_i}{\partial A_i}-\dfrac{\partial \mathcal{F}^\text{a}_i}{\partial A_i}\right]_{\bar{\bm{A}}^\text{r}_{l}}\right]\right]
 \text{det}\left(  \bm{\Gamma^\text{s}}(\bar{\bm{A}}^\text{r}_{l})\right)^{-1/2}  \\
&&\times\left.   \exp \left[\alpha^2\epsilon k_\text{B}T\left[\dfrac{\partial^2 \Gamma^{s}_{ij}}{\partial A_i\partial A_j} -\dfrac{\partial \Gamma^{s}_{ik}}{\partial A_k}({\Gamma^{s}}^{-1})_{ij} \dfrac{\partial \Gamma^{s}_{jp}}{\partial A_p}\right]_{\bar{\bm{A}}^\text{r}_{l}} +\mathcal{O}(\epsilon^{3/2})\right]\right\rbrace p_0(\bm{s}\circ\bm{A}_N).
\end{eqnarray}
}
\section{\hsn{Calculation of the ratio between the probability densities of a trajectory and its time-reversed trajectory}}\label{pratiosub} 
\hsn{Using relations~\eqref{Geq} \&~\eqref{Geq1}, expanding $N_{ij}(\bar{\bm{A}}^\text{f}_l)$ and $N_{ij}(\bar{\bm{A}}^\text{r}_l)$ around \hsn{$\bm{A}=\bar{\bm{A}}_l\equiv(\bm{A}_l+\bm{A}_{l-1})/2$}:
\begin{eqnarray}\label{Neqs}
N_{ij}(\bar{\bm{A}}^\text{f}_l)&=&N_{ij}(\bar{\bm{A}}_l)+\dfrac{2\alpha-1}{2}\left[ \dfrac{\partial N_{ij}}{\partial A_k}\right]_{\bar{\bm{A}}_l} dA_k(l)+\dfrac{1}{2}\left( \dfrac{2\alpha-1}{2}\right) ^2\left[ \dfrac{\partial^2 N_{ij}}{\partial A_k\partial A_m}\right]_{\bar{\bm{A}}_l} dA_k(l)dA_m(l)+\mathcal{O}(\epsilon^{3/2}),\\
N_{ij}(\bar{\bm{A}}^\text{r}_l)&=&N_{ij}(\bar{\bm{A}}_l)-\dfrac{2\alpha-1}{2}\left[ \dfrac{\partial N_{ij}}{\partial A_k}\right]_{\bar{\bm{A}}_l} dA_k(l)+\dfrac{1}{2}\left( \dfrac{2\alpha-1}{2}\right) ^2\left[ \dfrac{\partial^2 N_{ij}}{\partial A_k\partial A_m}\right]_{\bar{\bm{A}}_l} dA_k(l)dA_m(l)+\mathcal{O}(\epsilon^{3/2}).
\end{eqnarray}
Then, using the relation $N_{ik}N_{jk}=2 k_\text{B}T \Gamma^\text{s}_{ij}$ and Eq.~\eqref{logprop}, we obtain
\begin{eqnarray}\label{key}
\nonumber
\dfrac{ \text{det}\left(  \bm{\Gamma^\text{s}}(\bar{\bm{A}}^\text{f}_{l})\right)^{-1/2} }{ \text{det}\left(  \bm{\Gamma^\text{s}}(\bar{\bm{A}}^\text{r}_{l})\right)^{-1/2} }&=& \dfrac{\left|\text{det}\left(  \bm{N}^{-1}(\bar{\bm{A}}^\text{f}_l)\right) \right|\ }{\left|\text{det}\left(  \bm{N}^{-1}(\bar{\bm{A}}^\text{r}_{l})\right) \right|} \\\nonumber&=&\exp\left[ -(2\alpha-1)\left[\hsn{(N^{-1})_{jm}} \dfrac{\partial N_{mj}}{\partial A_k}\right]_{\bar{\bm{A}}_l} dA_k(l)+\mathcal{O}(\epsilon^{3/2})\right] \\
&=&\exp\left[-\dfrac{2\alpha-1}{2}\left[({\Gamma^{s}}^{-1})_{jm} \dfrac{\partial (\Gamma^{s})_{mj}}{\partial A_k}\right]_{\bar{\bm{A}}_l} dA_k(l)+\mathcal{O}(\epsilon^{3/2})\right] .
\end{eqnarray}
In $\epsilon \to 0$ limit, \hsfn{dividing} ~\eqref{AppendP222} by ~\eqref{AppendP223} first, and using the above equation and the relations given in Appendix~\ref{indent}, we get
\begin{eqnarray}\label{Pratiosup}
\nonumber
\dfrac{P}{P_r}&=&\dfrac{p_0(\bm{A}_0)}{p_0(\bm{s}\circ\bm{A}_N)}\exp\left[-\dfrac{1}{k_\text{B}T} \sum^N_{l=1}\left( \mathcal{H}(\bm{A}_l)-\mathcal{H}(\bm{A}_{l-1} )\right)+\mathcal{O}(\epsilon^{3/2} )\right]\\
&=&\dfrac{p_0(\bm{A}(0))}{p_0(\bm{s}\circ\bm{A}(\tau))}\exp\left[-\dfrac{1}{k_\text{B}T} \left( \mathcal{H}(\bm{A}(\tau))-\mathcal{H}(\bm{A}(0) )\right) \right],
\end{eqnarray}
where $\bm{A}(0)\equiv\bm{A}_0$ and $\bm{A}(\tau)\equiv\bm{A}_N$.
}
\section{Various relations needed for the calculation in subsection~\ref{ratio}}\label{indent}
Recalling Eq.~\eqref{geldisc} of the main text
\begin{equation}\label{D1}
dA_i(l)=\epsilon\mathcal{F}_i(\bar{\bm{A}}^\text{f}_l)+\sqrt{\epsilon}N_{ij}(\bar{\bm{A}}^\text{f}_l)\xi_j^{l}.
\end{equation}
Note that the lowest order term in $dA_i(l)$ is a $\epsilon^{1/2}$-term. Let us consider a function $G(\bm{A})$; expanding $G(\bar{\bm{A}}^\text{f}_l)$ \hsn{and $G(\bar{\bm{A}}^\text{r}_l)$} around \hsn{$\bm{A}=\bar{\bm{A}}_l\equiv(\bm{A}_l+\bm{A}_{l-1})/2$}:
\begin{eqnarray}\label{Geq}
\nonumber
G(\bar{\bm{A}}^\text{f}_l)&=&G(\bar{\bm{A}}_l+\dfrac{2\alpha-1}{2}d\bm{A}_l)\\
&=&G(\bar{\bm{A}}_l)+\dfrac{2\alpha-1}{2}\left[ \dfrac{\partial G}{\partial A_k}\right]_{\bar{\bm{A}}_l} dA_k(l)+\dfrac{1}{2}\left( \dfrac{2\alpha-1}{2}\right) ^2\left[ \dfrac{\partial^2 G}{\partial A_k\partial A_m}\right]_{\bar{\bm{A}}_l} dA_k(l)dA_m(l)+\mathcal{O}(\epsilon^{3/2})
\end{eqnarray}
\hsn{and
\begin{eqnarray}\label{Geq1}
\nonumber
G(\bar{\bm{A}}^\text{r}_l)&=&G(\bar{\bm{A}}_l-\dfrac{2\alpha-1}{2}d\bm{A}_l)\\
&=&G(\bar{\bm{A}}_l)-\dfrac{2\alpha-1}{2}\left[ \dfrac{\partial G}{\partial A_k}\right]_{\bar{\bm{A}}_l} dA_k(l)+\dfrac{1}{2}\left( \dfrac{2\alpha-1}{2}\right) ^2\left[ \dfrac{\partial^2 G}{\partial A_k\partial A_m}\right]_{\bar{\bm{A}}_l} dA_k(l)dA_m(l)+\mathcal{O}(\epsilon^{3/2}),
\end{eqnarray}
where} $d\bm{A}_l=\bm{A}_l-\bm{A}_{l-1}$. Then
\begin{eqnarray}\label{D3}
\nonumber
dA_i(l)dA_j(l)G(\bar{\bm{A}}^\text{f}_l)&=&dA_i(l)dA_j(l)G(\bar{\bm{A}}_l)+\dfrac{2\alpha-1}{2}\left[ \dfrac{\partial G}{\partial A_k}\right]_{\bar{\bm{A}}_l} dA_i(l)dA_j(l)dA_k(l)\\
&&+\dfrac{1}{2}\left( \dfrac{2\alpha-1}{2}\right) ^2\left[ \dfrac{\partial^2 G}{\partial A_k\partial A_m}\right]_{\bar{\bm{A}}_l} dA_i(l)dA_j(l)dA_k(l)dA_m(l)+\mathcal{O}(\epsilon^{5/2}).
\end{eqnarray}
From Eq.~\eqref{D1},
\begin{equation}\label{dadada0}
dA_i(l)dA_j(l)dA_k(l)=\xi_p^{l}\xi_q^{l}\xi_r^{l}N_{ip}N_{jq}N_{kr}\epsilon^{3/2}+(\xi_p^{l}\xi_q^{l}N_{ip}N_{jq}\mathcal{F}_k+\xi_p^{l}\xi_r^{l}N_{ip}N_{kr}\mathcal{F}_j+\xi_q^{l}\xi_r^{l}N_{jq}N_{kr}\mathcal{F}_i)\epsilon^2+\mathcal{O}(\epsilon^{5/2}).
\end{equation}
\hsn{Our final expressions will be written in integral form, and since $\xi_j(t)$ is the time derivative of a Wiener process, we can write:}
\begin{eqnarray}\label{key}
\label{D66}
\xi_p^{l}\xi_q^{l}&\equiv&\delta_{pq}\\
\xi_p^{l}\xi_q^{l}\xi_r^{l}&\equiv&\delta_{pq}\xi_r^{l}+\delta_{pr}\xi_q^{l}+\delta_{rq}\xi_p^{l}\\
\xi_p^{l}\xi_q^{l}\xi_r^{l}\xi_o^{l}&\equiv&\delta_{pq}\delta_{ro}+\delta_{pr}\delta_{qo}+\delta_{qr}\delta_{po}.
\end{eqnarray}
Using the above relations, Eq.~\eqref{D1} \& Eq.~\eqref{nn} ($N_{ik}N_{jk}=2 k_\text{B}T \Gamma^\text{s}_{ij}$), Eq.~\eqref{dadada0} can be written as

\begin{equation}\label{D6}
dA_i(l)dA_j(l)dA_k(l)=2 k_\text{B}T (\Gamma^\text{s}_{ij}  dA_k(l)+\Gamma^\text{s}_{jk} dA_i(l)+\Gamma^\text{s}_{ki}  dA_j(l))\epsilon+\mathcal{O}(\epsilon^{5/2}).
\end{equation}
Similarly,
\begin{eqnarray}\label{D8}
\nonumber
dA_i(l)dA_j(l)dA_k(l)dA_m(l)&=&\xi_p^{l}\xi_q^{l}\xi_r^{l}\xi_o^{l}N_{ip}N_{jq}N_{kr}N_{mo}\epsilon^2+\mathcal{O}(\epsilon^{5/2})\\
&= &(2k_\text{B}T)^2 (\Gamma^\text{s}_{ij}\Gamma^\text{s}_{km}+\Gamma^\text{s}_{ik}\Gamma^\text{s}_{jm}++\Gamma^\text{s}_{im}\Gamma^\text{s}_{jk})\epsilon^2+\mathcal{O}(\epsilon^{5/2})
\end{eqnarray}
Substituting~\eqref{D6} and ~\eqref{D8} into Eq.~\eqref{D3}, we obtain
\begin{eqnarray}\label{D9}
\nonumber
dA_i(l)dA_j(l)G(\bar{\bm{A}}^\text{f}_l)&=&dA_i(l)dA_j(l)G(\bar{\bm{A}}_l)+(2\alpha-1)k_\text{B}T\left[ \dfrac{\partial G}{\partial A_k}\right]_{\bar{\bm{A}}_l} (\Gamma^\text{s}_{ij}(\bar{\bm{A}}_l)  dA_k(l)+\Gamma^\text{s}_{jk}(\bar{\bm{A}}_l) dA_i(l)+\Gamma^\text{s}_{ki}(\bar{\bm{A}}_l)  dA_j(l))\epsilon\\
&&+\dfrac{1}{2}\left[ (2\alpha-1)k_\text{B}T\right] ^2\left[ \dfrac{\partial^2 G}{\partial A_k\partial A_m} (\Gamma^\text{s}_{ij}\Gamma^\text{s}_{km}+\Gamma^\text{s}_{ik}\Gamma^\text{s}_{jm}++\Gamma^\text{s}_{im}\Gamma^\text{s}_{jk})\right]_{\bar{\bm{A}}_l}\epsilon^2+\mathcal{O}(\epsilon^{5/2})
\end{eqnarray}
\hsn{Similarly, from Eq.~\eqref{Geq1}, we readily obtain}
\begin{eqnarray}\label{D31}
\nonumber
dA_i(l)dA_j(l)G(\bar{\bm{A}}^\text{r}_l)&=&dA_i(l)dA_j(l)G(\bar{\bm{A}}_l)-(2\alpha-1)k_\text{B}T\left[ \dfrac{\partial G}{\partial A_k}\right]_{\bar{\bm{A}}_l} (\Gamma^\text{s}_{ij}  dA_k(l)+\Gamma^\text{s}_{jk} dA_i(l)+\Gamma^\text{s}_{ki}  dA_j(l))\epsilon\\
&&+\dfrac{1}{2}\left[ (2\alpha-1)k_\text{B}T\right] ^2\left[ \dfrac{\partial^2 G}{\partial A_k\partial A_m}\right]_{\bar{\bm{A}}_l} (\Gamma^\text{s}_{ij}\Gamma^\text{s}_{km}+\Gamma^\text{s}_{ik}\Gamma^\text{s}_{jm}++\Gamma^\text{s}_{im}\Gamma^\text{s}_{jk})\epsilon^2+\mathcal{O}(\epsilon^{5/2}).
\end{eqnarray}
Likewise, we can easily derive the following relations:
\begin{eqnarray}\label{key}
dA_i(l)G(\bar{\bm{A}}^\text{f}_l)&=&dA_i(l)G(\bar{\bm{A}}_l)+(2\alpha-1)k_\text{B}T\epsilon \left[ \Gamma^{s}_{ij} \dfrac{\partial G}{\partial A_j }\right]_{\bar{\bm{A}}_l} +\mathcal{O}(\epsilon^{3/2})\\
dA_i(l)G(\bar{\bm{A}}^\text{r}_l)&=&dA_i(l)G(\bar{\bm{A}}_l)-(2\alpha-1)k_\text{B}T\epsilon \left[ \Gamma^{s}_{ij} \dfrac{\partial G}{\partial A_j }\right]_{\bar{\bm{A}}_l} +\mathcal{O}(\epsilon^{3/2})\\
G(\bar{\bm{A}}^\text{f}_l)&=&G(\bar{\bm{A}}_l)+\mathcal{O}(\epsilon^{1/2})\\
G(\bar{\bm{A}}^\text{r}_l)&=&G(\bar{\bm{A}}_l)+\mathcal{O}(\epsilon^{1/2})\\
\left[ \dfrac{\partial G}{\partial A_i}\right]_{\bar{\bm{A}}_l}dA_i(l)&=&G(\bm{A}_l)-G(\bm{A}_{l-1})+\mathcal{O}(\epsilon^{3/2})
\end{eqnarray}
\end{widetext}

\section{The relation between entropy production rate and $\dot{s}$}\label{thermodynamics}
 The free energy of the system at time $t$ would be
\begin{eqnarray}\label{free0}
\nonumber
F(t)&=& \int \mathcal{H}(\bm{A})p_t(\bm{A})d\bm{A} - T\left[ -k_{\text{B}}\int p_t(\bm{A})\ln p_t(\bm{A})d\bm{A}\right]  \\
&=& \left\langle\left[ \mathcal{H}(\bm{A}) +k_{\text{B}}T \ln p_t(\bm{A})\right] \right\rangle,
\end{eqnarray}
where $\left\langle \right\rangle $ stands for the ensemble average and $p_t(\bm{A})$ is the probability distribution of $\bm{A}$ at time $t$. Let us define the free energy of a single trajectory of the system at time $t$ as
\begin{equation}\label{free1}
f(t)= \mathcal{H}(\bm{A}(t)) +k_{\text{B}}T \ln p_t(\bm{A}(t))
\end{equation}
Then it is straight forward to show that
\begin{eqnarray}\label{free2}
\dfrac{d{F}(t)}{dt}=\left\langle  \dfrac{df(t)}{dt} \right\rangle.
\end{eqnarray}

\subsection{For passive systems}
From Eq~\eqref{sdotpass}, we readily get
\begin{equation}\label{key}
\left\langle  \dfrac{df(t)}{dt}\right\rangle= -k_{\text{B}}T\left\langle \dot{s}(t)\right\rangle,
\end{equation} 
so from Eq.~\eqref{free2}
\begin{eqnarray}\label{free3}
\dfrac{d{F}(t)}{dt}=-k_{\text{B}}T\left\langle \dot{s}(t)\right\rangle.
\end{eqnarray}
Assuming that the system is always in metastable thermal equilibrium with the reservoir, the rate of change total entropy of system and reservoir would be 
\begin{equation}\label{entpass}
\dfrac{dS(t)}{dt}=-\dfrac{1}{T}\dfrac{d{F}(t)}{dt}=k_{\text{B}}\left\langle \dot{{s}}(t)\right\rangle.
\end{equation}
\subsection{For active systems}
For active systems, from Eq.~\eqref{rdotdri}, one can trivially prove that
\begin{equation}\label{key}
\left\langle  \dfrac{df(t)}{dt}\right\rangle= -k_{\text{B}}T\left\langle \dot{s}(t)\right\rangle  +\left\langle w(t)\right\rangle,
\end{equation}
where $\left\langle w(t)\right\rangle$ is average rate of the work perfomed by active forces.
Hence, from Eq.~\eqref{free2},
\begin{equation}\label{key}
\dfrac{d{F}(t)}{dt}=-k_{\text{B}}T\left\langle \dot{{s}}(t)\right\rangle + \left\langle w(t)\right\rangle.
\end{equation}
Therefore, the rate of total enetropy production of the system and the reservoir:
\begin{eqnarray}
\nonumber
\dfrac{dS(t)}{dt}&=&-\dfrac{1}{T}\dfrac{d{F}(t)}{dt}+\dfrac{1}{T}\left\langle w(t)\right\rangle\\
\label{entact}
&=&k_{\text{B}}\left\langle \dot{{s}}(t)\right\rangle.
\end{eqnarray}

\section{The dissipation function defined by Seifert \textit{et al.}~\cite{Seifert_prl_2005}}~\label{seif}
Seifert \textit{et al.}~\cite{Seifert_prl_2005} used the following form of the dissipation function:
\begin{equation}\label{key}
\mathcal{S}_\tau=\ln \left[ \dfrac{P}{P'_r}\right],
\end{equation}
where $P$ is the probability density of a trajectory between $t=0$ and $t=\tau$ which is given by Eq.~\eqref{AppendP22}, and  $P'_r$ is the probability density of the time-reversed trajectory, considering that the time-reserved trajectory starts at $t=\tau$, not at $t=0$. Thus, the epxression of $\mathcal{S}_\tau$ is readily obtained by replacing $p_0(\bm{s}\circ\bm{A}_N)$ with $p_t(\bm{s}\circ\bm{A}_N)$ in the expression of $\mathcal{R}_\tau$ (see Eq.~\eqref{dissf}), that is
\begin{equation}\label{key}
\mathcal{S}_\tau=\ln \dfrac{p_0(\bm{A}(0))}{p_\tau(\bm{s}\circ\bm{A}(\tau))}-\dfrac{1}{k_\text{B}T}\left[  \mathcal{H}(\bm{A}(\tau)) -\mathcal{H}(\bm{A}(0))\right]
\end{equation}
This dissipation function follows the fluctuation relation~\eqref{ft} in steady states only, not in general. However, as discussed by~\cite{Seifert_prl_2005}, it does always follow the integrated fluctuation relation~\eqref{ift} and thereofore $\left\langle \mathcal{S}_\tau\right\rangle \geq 0 $.

\section{Calculation of $\dot{s}$ for the active systems}\label{sdot}
Recalling equations of motion for the active systems
\begin{equation}\label{key}
\dfrac{dA_i}{dt}=\mathcal{F}_i+\mathcal{X}_i+N_{ij}\xi_j(t),
\end{equation}
where 
\begin{equation}\label{key}
\mathcal{F}_i\equiv-\Gamma_{ij}\dfrac{\partial \mathcal{H}}{\partial A_j}+k_\text{B}T\dfrac{\partial \Gamma_{ij}}{\partial A_j}-\alpha N_{lj}  \dfrac{\partial N_{ij}}{\partial A_l}
\end{equation}
and $\mathcal{X}_i$ is the active term. Writing \hsn{$\bm{\mathcal{X}}$ as $\bm{\mathcal{X}}=\bm{\mathcal{X}}^\text{s}+\bm{\mathcal{X}}^\text{a}$, where
\begin{eqnarray}
\label{Xs}
\bm{\mathcal{X}}^\text{s}(\bm{A})&=&\dfrac{1}{2}\left( \bm{\mathcal{X}}(\bm{A})+\bm{s}\circ\bm{\mathcal{X}}(\bm{s}\circ\bm{A})\right), \\
\label{Xa}
\bm{\mathcal{X}}^\text{a}(\bm{A})&=&\dfrac{1}{2}\left( \bm{\mathcal{X}}(\bm{A})-\bm{s}\circ\bm{\mathcal{X}}(\bm{s}\circ\bm{A})\right)
\end{eqnarray}	
follow the properties $\bm{\mathcal{X}}^\text{s}(\bm{s}\circ\bm{A})= \bm{s}\circ \bm{\mathcal{X}}^\text{s}(\bm{A})$ and $\bm{\mathcal{X}}^\text{a}(\bm{s}\circ\bm{A})= -\bm{s}\circ \bm{\mathcal{X}}^\text{a}(\bm{A})$.} 
The Fokker-Planck equation for the probability density $p_t(\bm{A})$ of  $\bm{A}$ reads
\begin{equation}\label{fpe}
\dfrac{\partial p_t(\bm{A})}{\partial t}=-\dfrac{\partial J_i(\bm{A},t) }{\partial A_i},
\end{equation}
where 
\begin{eqnarray}\label{JIap}
\nonumber
J_i(\bm{A},t)&=&\left( -{\Gamma^{\text{s}}}_{ij}(\bm{A})\dfrac{\partial \mathcal{H}(\bm{A})}{\partial A_j}+\mathcal{Y}^\text{a}_i(\bm{A}) +\mathcal{X}^\text{s}_j(\bm{A})\right) p_t(\bm{A})\\&&\qquad\qquad\qquad\qquad-k_\text{B}T\Gamma^{\text{s}}_{ij}(\bm{A})\dfrac{\partial p_t(\bm{A})}{dA_j}
\end{eqnarray}	
is the probability current~\cite{lau2007state} and
\begin{equation}\label{key}
\mathcal{Y}^\text{a}_i= \mathcal{X}^\text{a}_i +\dfrac{\partial \mathcal{H}}{\partial A_k}\Gamma^\text{a}_{ki}-k_\text{B}T\dfrac{\partial \Gamma^\text{a}_{ki}}{\partial A_k}.
\end{equation}
\hsn{Since $\bm{\mathcal{X}}^\text{a}(\bm{s}\circ\bm{A})= -\bm{s}\circ \bm{\mathcal{X}}^\text{a}(\bm{A})$, using the relation $\Gamma^\text{a}_{ij}=-\Gamma^\text{a}_{ij}s_is_j$ (Eq.~\eqref{Gammaapro}), it is easy to show that 
\begin{equation}\label{yasyn}
\bm{\mathcal{Y}}^\text{a}(\bm{s}\circ\bm{A})=-\bm{s}\circ\bm{\mathcal{Y}}^\text{a}(\bm{A}).
\end{equation}	
As Eq.~\eqref{fpe} has no term with $N_{ij}$, $p_t(\bm{A})$ would be independent of the choice of $N_{ij}$.}
Recalling Eq.~\eqref{disspumped}
\begin{equation}\label{disspumpedap}
\dot{s}(t)=w(t) -\left.\dfrac{d}{dt'}\left( \ln p_t(\bm{A}(t'))+\dfrac{1}{k_\text{B}T}\mathcal{H}(\bm{A}(t'))\right)\right|_{t'=t}.
\end{equation}
where 
\begin{eqnarray}\label{wtap}
\nonumber
&w(t)&= \dfrac{\partial \mathcal{H}(\bm{A}(t))}{\partial A_i}\mathcal{X}^\text{a}_i(\bm{A}(t))-k_\text{B}T\dfrac{\partial \mathcal{X}^\text{a}_i(\bm{A}(t))}{\partial A_i} \\&+&\hsn{({\Gamma^{\text{s}}}^{-1})_{ij}}(\bm{A}(t)) \mathcal{X}^\text{s}_j(\bm{A}(t))\left[\dfrac{dA_i}{dt}-\mathcal{Y}^\text{a}_i(\bm{A}(t))\right].
\end{eqnarray}
Using Eqs.~~\eqref{fpe} \& \eqref{JIap}, one can write Eq.~\eqref{disspumped} in the following form:
\begin{eqnarray}
\nonumber
\dot{{s}}(t)&=&\dfrac{1}{k_\text{B}T}\left(\dfrac{J_i(\bm{A},t)}{p_t(\bm{A})}-\mathcal{Y}^\text{a}_i(\bm{A}) \right)\hsn{({\Gamma^{\text{s}}}^{-1})_{ij}}(\bm{A})\\
\label{stnew}
&&\times \left( \dfrac{J_j(\bm{A},t)}{p_t(\bm{A})}-\mathcal{Y}^\text{a}_j(\bm{A})\right)+R_s+R_0,
\end{eqnarray}
where 
\begin{equation}\label{key}
R_s=-\dfrac{1}{p_t(\bm{A})}\dfrac{\partial}{\partial A_i}\left( p_t(\bm{A})\mathcal{Y}^\text{a}_i(\bm{A}) \right)
\end{equation}
and 
\begin{eqnarray}
\nonumber
&R_0&=\left(\dfrac{dA_i}{dt}-\dfrac{J_i(\bm{A},t)}{p_t(\bm{A})} \right) \left[ \dfrac{1}{p_t(\bm{A})}\dfrac{\partial p_t(\bm{A})}{\partial A_i}\right. \\&&\left.-  \dfrac{1}{k_\text{B}T}\left( \hsn{({\Gamma^{\text{s}}}^{-1})_{ij}}(\bm{A}) \mathcal{X}^\text{s}_j(\bm{A})-\dfrac{\partial \mathcal{H}(\bm{A})}{\partial A_j}\right)\right].
\end{eqnarray}
\hsn{The average of $R_s$ reads
\begin{eqnarray}\label{Rs}
\nonumber
\left\langle R_s\right\rangle &=&-\int   \dfrac{1}{p_t(\bm{A})}\dfrac{\partial}{\partial A_i}\left( p_t(\bm{A})\mathcal{Y}^\text{a}_i(\bm{A}) \right)p_t(\bm{A})d^nA\\
&=&-\int   \dfrac{\partial}{\partial A_i}\left( p_t(\bm{A})\mathcal{Y}^\text{a}_i(\bm{A}) \right)d^nA
\end{eqnarray}	
The above expression can be written as a surface integral with the integrand $\bm{I}_\text{s}=-p_t(\bm{A})\bm{\mathcal{Y}}^\text{a}(\bm{A})$.  If the system is periodic in $A_i$ (e.g., $A_i$ is an angle), the surface integral is already zero. If $A_i$ lies in the infinite interval $(-\infty,\infty)$,   in $A_i\to\pm\infty$ limit, $p_t(\bm{A})\to0$, given that $\bm{A}$ are physical variables. Assuming that $||\bm{I}_\text{s}||$ converges faster than $||\bm{A}||^{1-n}$,  the surface integral is again zero. For  $p_t(\bm{s}\circ\bm{A})=p_t(\bm{A})$ case, $\left\langle R_s\right\rangle$ is always zero as follows:  setting $\bm{A}=\bm{s}\circ\bm{A}'$ gives $d^nA=d^nA'$, then using Eq.~\eqref{yasyn}, we get
\begin{eqnarray}\label{key}
\nonumber
\left\langle R_s\right\rangle &=&\int   \dfrac{\partial}{\partial A'_i}\left( p_t(\bm{A}')\mathcal{Y}^\text{a}_i(\bm{A}') \right)d^nA'\\
&=&-\left\langle R_s\right\rangle,
\end{eqnarray}
so $\left\langle R_s\right\rangle=0$.
 Since the ensemble average of $dA_i/dt$ for given $\bm{A}$ and $t$  is just $J_i(\bm{A},t)/p_t(\bm{A})$, $\left\langle R_o\right\rangle=0$.
 Therefore,} the ensemble average of Eq.~\eqref{stnew} is given by
\begin{eqnarray}\label{key}
\nonumber
\left\langle \dot{{s}}(t)\right\rangle&=&\dfrac{1}{k_\text{B}T}\left\langle \left(\dfrac{J_i(\bm{A},t)}{p_t(\bm{A})}-\mathcal{Y}^\text{a}_i(\bm{A}) \right)\hsn{({\Gamma^{\text{s}}}^{-1})_{ij}}(\bm{A})\right. \\
&&\times \left. \left( \dfrac{J_j(\bm{A},t)}{p_t(\bm{A})}-\mathcal{Y}^\text{a}_j(\bm{A})\right)\right\rangle.
\end{eqnarray}
Similarly, $\left\langle w(t)\right\rangle $ given by Eq.~\eqref{wtap} can be written in the following form
\begin{eqnarray}\label{key}
\nonumber
\left\langle w(t)\right\rangle &=&\left\langle \left( \dfrac{dA_i}{dt}-\dfrac{\partial \mathcal{H}}{\partial A_k}\Gamma^\text{a}_{ki}+k_\text{B}T\dfrac{\partial \Gamma^\text{a}_{ki}}{\partial A_k}\right) \hsn{({\Gamma^{\text{s}}}^{-1})_{ij}} \mathcal{X}^\text{s}_j\right\rangle\\&&-\left\langle \left( \dfrac{dA_i}{dt}- \mathcal{Y}^\text{a}_i\right)\hsn{({\Gamma^{\text{s}}}^{-1})_{ij}} \mathcal{X}^\text{a}_j \right\rangle.
\end{eqnarray}
\hsn{Here, if $p_t(\bm{s}\circ\bm{A})\neq p_t(\bm{A})$, we must assume that $||p_t(\bm{A})\bm{\mathcal{X}}^\text{a}(\bm{A})||$ converges faster than $||\bm{A}||^{1-n}$.}
\section{Stationary solution of the Fokker-Planck equation associated with Eq.~\eqref{gle2}}\label{fps}
\hsn{The Fokker-Planck equation for the probability distribution $p_t(\bm{A})$ of the solution of Eq~\eqref{gle2} is given by
\begin{equation}\label{fpep}
\dfrac{\partial p_t(\bm{A})}{\partial t}=-\dfrac{\partial J_i(\bm{A},t) }{\partial A_i};
\end{equation}
the expression of the probability current $J_i(\bm{A},t)$ reads~\cite{lau2007state} 
\begin{eqnarray}\label{JIapp1}
\nonumber
J_i(\bm{A},t)&=&\left( -{\Gamma^{\text{s}}}_{ij}(\bm{A})\dfrac{\partial \mathcal{H}(\bm{A})}{\partial A_j} +\dfrac{\partial \mathcal{H}}{\partial A_k}\Gamma^\text{a}_{ki}-k_\text{B}T\dfrac{\partial \Gamma^\text{a}_{ki}}{\partial A_k}\right) \\&&\times p_t(\bm{A})-k_\text{B}T\Gamma^{\text{s}}_{ij}(\bm{A})\dfrac{\partial p_t(\bm{A})}{dA_j}.
\end{eqnarray}	
Undoubtedly, the dynamics of $ p_t(\bm{A})$ is independent of the choice of $N_{ij}$ and $\alpha$.
Since  
\begin{equation}\label{key}
\nonumber
\dfrac{\partial \mathcal{H}}{\partial A_k}\Gamma^\text{a}_{ki}-k_\text{B}T\dfrac{\partial \Gamma^\text{a}_{ki}}{\partial A_k}
\end{equation}
is the Poisson bracket term, the stationary solution of the above equation is given by~\cite{Lubensky_book}
\begin{equation}\label{psseqpdf}
p_\text{s}(\bm{A})=\dfrac{1}{\mathcal{Z}}\exp \left[ -\dfrac{\mathcal{H}(\bm{A})}{k_\text{B}T}\right],
\end{equation}
where $Z$ is the normalizing constant. }
\section{Dependence of $\left\langle \mathcal{R}_\tau\right\rangle $ on $\alpha$ for the passive systems}\label{rtaui}
\hsn{ For the passive systems,  $\mathcal{R}_\tau$ depends only on the initial and final states of the system $\bm{A}_0$ and $\bm{A}_{\tau}$, so its average can be calculated using the formula~\cite{doibookpolymer}
\begin{equation}\label{key}
\left\langle \mathcal{R}_\tau\right\rangle =\int\mathcal{R}_\tau p_0(\bm{A}_0)G(\bm{A}_0,\bm{A}_{\tau};\tau)d^nA_0d^nA_{\tau},
\end{equation}
where $p_0(\bm{A})$ is the probability distribution of $\bm{A}$ at $t=0$, and $G(\bm{A},\bm{A}';\tau)$ is the probability distribution of state $\bm{A}'$ at $t=\tau$ given that the system was in the state $\bm{A}$ at $t=0$; it is the solution of Eq.~\eqref{fpep} with the initial condition  $G(\bm{A},\bm{A}';\tau=0)=\bm{\delta}(\bm{A}-\bm{A}')$. As the solution of Eq.~\eqref{fpep} is independent of $\alpha$, $\left\langle \mathcal{R}_\tau\right\rangle$ would be constant in $\alpha$.}

%


\begin{thebibliography}{35}%
	\makeatletter
	\providecommand \@ifxundefined [1]{%
		\@ifx{#1\undefined}
	}%
	\providecommand \@ifnum [1]{%
		\ifnum #1\expandafter \@firstoftwo
		\else \expandafter \@secondoftwo
		\fi
	}%
	\providecommand \@ifx [1]{%
		\ifx #1\expandafter \@firstoftwo
		\else \expandafter \@secondoftwo
		\fi
	}%
	\providecommand \natexlab [1]{#1}%
	\providecommand \enquote  [1]{``#1''}%
	\providecommand \bibnamefont  [1]{#1}%
	\providecommand \bibfnamefont [1]{#1}%
	\providecommand \citenamefont [1]{#1}%
	\providecommand \href@noop [0]{\@secondoftwo}%
	\providecommand \href [0]{\begingroup \@sanitize@url \@href}%
	\providecommand \@href[1]{\@@startlink{#1}\@@href}%
	\providecommand \@@href[1]{\endgroup#1\@@endlink}%
	\providecommand \@sanitize@url [0]{\catcode `\\12\catcode `\$12\catcode
		`\&12\catcode `\#12\catcode `\^12\catcode `\_12\catcode `\%12\relax}%
	\providecommand \@@startlink[1]{}%
	\providecommand \@@endlink[0]{}%
	\providecommand \url  [0]{\begingroup\@sanitize@url \@url }%
	\providecommand \@url [1]{\endgroup\@href {#1}{\urlprefix }}%
	\providecommand \urlprefix  [0]{URL }%
	\providecommand \Eprint [0]{\href }%
	\providecommand \doibase [0]{https://doi.org/}%
	\providecommand \selectlanguage [0]{\@gobble}%
	\providecommand \bibinfo  [0]{\@secondoftwo}%
	\providecommand \bibfield  [0]{\@secondoftwo}%
	\providecommand \translation [1]{[#1]}%
	\providecommand \BibitemOpen [0]{}%
	\providecommand \bibitemStop [0]{}%
	\providecommand \bibitemNoStop [0]{.\EOS\space}%
	\providecommand \EOS [0]{\spacefactor3000\relax}%
	\providecommand \BibitemShut  [1]{\csname bibitem#1\endcsname}%
	\let\auto@bib@innerbib\@empty
	\bibitem [{\citenamefont {Evans}\ \emph {et~al.}(1993)\citenamefont {Evans},
		\citenamefont {Cohen},\ and\ \citenamefont {Morriss}}]{evans_first_prl}%
	\BibitemOpen
	\bibfield  {author} {\bibinfo {author} {\bibfnamefont {D.~J.}\ \bibnamefont
			{Evans}}, \bibinfo {author} {\bibfnamefont {E.~G.~D.}\ \bibnamefont
			{Cohen}},\ and\ \bibinfo {author} {\bibfnamefont {G.~P.}\ \bibnamefont
			{Morriss}},\ }\bibfield  {title} {\bibinfo {title} {Probability of second law
			violations in shearing steady states},\ }\href
	{https://doi.org/10.1103/PhysRevLett.71.2401} {\bibfield  {journal} {\bibinfo
			{journal} {Phys. Rev. Lett.}\ }\textbf {\bibinfo {volume} {71}},\ \bibinfo
		{pages} {2401} (\bibinfo {year} {1993})}\BibitemShut {NoStop}%
	\bibitem [{\citenamefont {Evans}\ and\ \citenamefont
		{Searles}(2002)}]{evans_ad_phy}%
	\BibitemOpen
	\bibfield  {author} {\bibinfo {author} {\bibfnamefont {D.~J.}\ \bibnamefont
			{Evans}}\ and\ \bibinfo {author} {\bibfnamefont {D.~J.}\ \bibnamefont
			{Searles}},\ }\bibfield  {title} {\bibinfo {title} {The fluctuation
			theorem},\ }\href {https://doi.org/10.1080/00018730210155133} {\bibfield
		{journal} {\bibinfo  {journal} {Advances in Physics}\ }\textbf {\bibinfo
			{volume} {51}},\ \bibinfo {pages} {1529} (\bibinfo {year}
		{2002})}\BibitemShut {NoStop}%
	\bibitem [{\citenamefont {Searles}\ and\ \citenamefont
		{Evans}(2004)}]{Searles2004}%
	\BibitemOpen
	\bibfield  {author} {\bibinfo {author} {\bibfnamefont {D.~J.}\ \bibnamefont
			{Searles}}\ and\ \bibinfo {author} {\bibfnamefont {D.~J.}\ \bibnamefont
			{Evans}},\ }\bibfield  {title} {\bibinfo {title} {Fluctuations relations for
			nonequilibrium systems},\ }\href {http://dx.doi.org/10.1071/CH04115}
	{\bibfield  {journal} {\bibinfo  {journal} {Australian Journal of Chemistry}\
		}\textbf {\bibinfo {volume} {57}},\ \bibinfo {pages} {1119} (\bibinfo {year}
		{2004})}\BibitemShut {NoStop}%
	\bibitem [{\citenamefont {Liphardt}\ \emph {et~al.}(2002)\citenamefont
		{Liphardt}, \citenamefont {Dumont}, \citenamefont {Smith}, \citenamefont
		{Tinoco},\ and\ \citenamefont {Bustamante}}]{Jarz_equlity_scince_2002}%
	\BibitemOpen
	\bibfield  {author} {\bibinfo {author} {\bibfnamefont {J.}~\bibnamefont
			{Liphardt}}, \bibinfo {author} {\bibfnamefont {S.}~\bibnamefont {Dumont}},
		\bibinfo {author} {\bibfnamefont {S.~B.}\ \bibnamefont {Smith}}, \bibinfo
		{author} {\bibfnamefont {I.}~\bibnamefont {Tinoco}},\ and\ \bibinfo {author}
		{\bibfnamefont {C.}~\bibnamefont {Bustamante}},\ }\bibfield  {title}
	{\bibinfo {title} {Equilibrium information from nonequilibrium measurements
			in an experimental test of jarzynski's equality},\ }\href
	{https://doi.org/10.1126/science.1071152} {\ \textbf {\bibinfo {volume}
			{296}},\ \bibinfo {pages} {1832} (\bibinfo {year} {2002})}\BibitemShut
	{NoStop}%
	\bibitem [{\citenamefont {Seifert}(2005)}]{Seifert_prl_2005}%
	\BibitemOpen
	\bibfield  {author} {\bibinfo {author} {\bibfnamefont {U.}~\bibnamefont
			{Seifert}},\ }\bibfield  {title} {\bibinfo {title} {Entropy production along
			a stochastic trajectory and an integral fluctuation theorem},\ }\href
	{https://doi.org/10.1103/PhysRevLett.95.040602} {\bibfield  {journal}
		{\bibinfo  {journal} {Phys. Rev. Lett.}\ }\textbf {\bibinfo {volume} {95}},\
		\bibinfo {pages} {040602} (\bibinfo {year} {2005})}\BibitemShut {NoStop}%
	\bibitem [{\citenamefont {Blickle}\ \emph {et~al.}(2006)\citenamefont
		{Blickle}, \citenamefont {Speck}, \citenamefont {Helden}, \citenamefont
		{Seifert},\ and\ \citenamefont {Bechinger}}]{seifert2006prl}%
	\BibitemOpen
	\bibfield  {author} {\bibinfo {author} {\bibfnamefont {V.}~\bibnamefont
			{Blickle}}, \bibinfo {author} {\bibfnamefont {T.}~\bibnamefont {Speck}},
		\bibinfo {author} {\bibfnamefont {L.}~\bibnamefont {Helden}}, \bibinfo
		{author} {\bibfnamefont {U.}~\bibnamefont {Seifert}},\ and\ \bibinfo {author}
		{\bibfnamefont {C.}~\bibnamefont {Bechinger}},\ }\bibfield  {title} {\bibinfo
		{title} {Thermodynamics of a colloidal particle in a time-dependent
			nonharmonic potential},\ }\href
	{https://doi.org/10.1103/PhysRevLett.96.070603} {\bibfield  {journal}
		{\bibinfo  {journal} {Phys. Rev. Lett.}\ }\textbf {\bibinfo {volume} {96}},\
		\bibinfo {pages} {070603} (\bibinfo {year} {2006})}\BibitemShut {NoStop}%
	\bibitem [{\citenamefont {Seifert}(2012)}]{seifert2012stochastic}%
	\BibitemOpen
	\bibfield  {author} {\bibinfo {author} {\bibfnamefont {U.}~\bibnamefont
			{Seifert}},\ }\bibfield  {title} {\bibinfo {title} {Stochastic
			thermodynamics, fluctuation theorems and molecular machines},\ }\href@noop {}
	{\bibfield  {journal} {\bibinfo  {journal} {Reports on progress in physics}\
		}\textbf {\bibinfo {volume} {75}},\ \bibinfo {pages} {126001} (\bibinfo
		{year} {2012})}\BibitemShut {NoStop}%
	\bibitem [{\citenamefont {Gallavotti}(2008)}]{gallavotti_epjb_2008}%
	\BibitemOpen
	\bibfield  {author} {\bibinfo {author} {\bibfnamefont {G.}~\bibnamefont
			{Gallavotti}},\ }\bibfield  {title} {\bibinfo {title} {Heat and fluctuations
			from order to chaos},\ }\href {https://doi.org/10.1140/epjb/e2008-00041-1}
	{\bibfield  {journal} {\bibinfo  {journal} {The European Physical Journal B}\
		}\textbf {\bibinfo {volume} {61}},\ \bibinfo {pages} {1} (\bibinfo {year}
		{2008})}\BibitemShut {NoStop}%
	\bibitem [{\citenamefont {Jarzynski}(2007)}]{jarzynski2007comparison}%
	\BibitemOpen
	\bibfield  {author} {\bibinfo {author} {\bibfnamefont {C.}~\bibnamefont
			{Jarzynski}},\ }\bibfield  {title} {\bibinfo {title} {Comparison of
			far-from-equilibrium work relations},\ }\href@noop {} {\bibfield  {journal}
		{\bibinfo  {journal} {Comptes Rendus Physique}\ }\textbf {\bibinfo {volume}
			{8}},\ \bibinfo {pages} {495} (\bibinfo {year} {2007})}\BibitemShut {NoStop}%
	\bibitem [{\citenamefont {Hurtado}\ \emph {et~al.}(2011)\citenamefont
		{Hurtado}, \citenamefont {P\`erez-Espigares}, \citenamefont {del Pozo},\ and\
		\citenamefont {Garrido}}]{ifr_pnas_2011}%
	\BibitemOpen
	\bibfield  {author} {\bibinfo {author} {\bibfnamefont {P.~I.}\ \bibnamefont
			{Hurtado}}, \bibinfo {author} {\bibfnamefont {C.}~\bibnamefont
			{P\`erez-Espigares}}, \bibinfo {author} {\bibfnamefont {J.~J.}\ \bibnamefont
			{del Pozo}},\ and\ \bibinfo {author} {\bibfnamefont {P.~L.}\ \bibnamefont
			{Garrido}},\ }\bibfield  {title} {\bibinfo {title} {Symmetries in
			fluctuations far from equilibrium},\ }\href
	{http://www.pnas.org/content/early/2011/04/12/1013209108.abstract} {\bibfield
		{journal} {\bibinfo  {journal} {Proceedings of the National Academy of
				Sciences}\ } (\bibinfo {year} {2011})}\BibitemShut {NoStop}%
	\bibitem [{\citenamefont {Villavicencio-Sanchez}\ \emph
		{et~al.}(2014)\citenamefont {Villavicencio-Sanchez}, \citenamefont {Harris},\
		and\ \citenamefont {Touchette}}]{AFR_EPL_2014}%
	\BibitemOpen
	\bibfield  {author} {\bibinfo {author} {\bibfnamefont {R.}~\bibnamefont
			{Villavicencio-Sanchez}}, \bibinfo {author} {\bibfnamefont {R.~J.}\
			\bibnamefont {Harris}},\ and\ \bibinfo {author} {\bibfnamefont
			{H.}~\bibnamefont {Touchette}},\ }\bibfield  {title} {\bibinfo {title}
		{Fluctuation relations for anisotropic systems},\ }\href
	{http://stacks.iop.org/0295-5075/105/i=3/a=30009} {\bibfield  {journal}
		{\bibinfo  {journal} {EPL}\ }\textbf {\bibinfo {volume} {105}},\ \bibinfo
		{pages} {30009} (\bibinfo {year} {2014})}\BibitemShut {NoStop}%
	\bibitem [{\citenamefont {Soni}(2019)}]{soni2019flocks}%
	\BibitemOpen
	\bibfield  {author} {\bibinfo {author} {\bibfnamefont {H.}~\bibnamefont
			{Soni}},\ }\emph {\bibinfo {title} {Flocks, Flow and Fluctuations in
			Inanimate Matter: Simulations and Theory}},\ \href@noop {} {Ph.D. thesis}
	(\bibinfo {year} {2019})\BibitemShut {NoStop}%
	\bibitem [{\citenamefont {Wang}\ \emph {et~al.}(2002)\citenamefont {Wang},
		\citenamefont {Sevick}, \citenamefont {Mittag}, \citenamefont {Searles},\
		and\ \citenamefont {Evans}}]{op_tweez_exp_prl_2002}%
	\BibitemOpen
	\bibfield  {author} {\bibinfo {author} {\bibfnamefont {G.~M.}\ \bibnamefont
			{Wang}}, \bibinfo {author} {\bibfnamefont {E.~M.}\ \bibnamefont {Sevick}},
		\bibinfo {author} {\bibfnamefont {E.}~\bibnamefont {Mittag}}, \bibinfo
		{author} {\bibfnamefont {D.~J.}\ \bibnamefont {Searles}},\ and\ \bibinfo
		{author} {\bibfnamefont {D.~J.}\ \bibnamefont {Evans}},\ }\bibfield  {title}
	{\bibinfo {title} {Experimental demonstration of violations of the second law
			of thermodynamics for small systems and short time scales},\ }\href
	{https://doi.org/10.1103/PhysRevLett.89.050601} {\bibfield  {journal}
		{\bibinfo  {journal} {Phys. Rev. Lett.}\ }\textbf {\bibinfo {volume} {89}},\
		\bibinfo {pages} {050601} (\bibinfo {year} {2002})}\BibitemShut {NoStop}%
	\bibitem [{\citenamefont {Collin}\ \emph {et~al.}(2005)\citenamefont {Collin},
		\citenamefont {Ritort}, \citenamefont {Jarzynski}, \citenamefont {Smith},
		\citenamefont {Tinoco},\ and\ \citenamefont
		{Bustamante}}]{crook_dna_nature_2005}%
	\BibitemOpen
	\bibfield  {author} {\bibinfo {author} {\bibfnamefont {D.}~\bibnamefont
			{Collin}}, \bibinfo {author} {\bibfnamefont {F.}~\bibnamefont {Ritort}},
		\bibinfo {author} {\bibfnamefont {C.}~\bibnamefont {Jarzynski}}, \bibinfo
		{author} {\bibfnamefont {S.~B.}\ \bibnamefont {Smith}}, \bibinfo {author}
		{\bibfnamefont {I.}~\bibnamefont {Tinoco}},\ and\ \bibinfo {author}
		{\bibfnamefont {C.}~\bibnamefont {Bustamante}},\ }\bibfield  {title}
	{\bibinfo {title} {Verification of the crooks fluctuation theorem and
			recovery of rna folding free energies},\ }\href
	{https://doi.org/10.1038/nature04061} {\bibfield  {journal} {\bibinfo
			{journal} {Nature}\ }\textbf {\bibinfo {volume} {437}},\ \bibinfo {pages}
		{231} (\bibinfo {year} {2005})}\BibitemShut {NoStop}%
	\bibitem [{\citenamefont {Carberry}\ \emph {et~al.}(2004)\citenamefont
		{Carberry}, \citenamefont {Reid}, \citenamefont {Wang}, \citenamefont
		{Sevick}, \citenamefont {Searles},\ and\ \citenamefont
		{Evans}}]{op_tweez_exp_prl_2004}%
	\BibitemOpen
	\bibfield  {author} {\bibinfo {author} {\bibfnamefont {D.~M.}\ \bibnamefont
			{Carberry}}, \bibinfo {author} {\bibfnamefont {J.~C.}\ \bibnamefont {Reid}},
		\bibinfo {author} {\bibfnamefont {G.~M.}\ \bibnamefont {Wang}}, \bibinfo
		{author} {\bibfnamefont {E.~M.}\ \bibnamefont {Sevick}}, \bibinfo {author}
		{\bibfnamefont {D.~J.}\ \bibnamefont {Searles}},\ and\ \bibinfo {author}
		{\bibfnamefont {D.~J.}\ \bibnamefont {Evans}},\ }\bibfield  {title} {\bibinfo
		{title} {Fluctuations and irreversibility: An experimental demonstration of a
			second-law-like theorem using a colloidal particle held in an optical trap},\
	}\href {https://doi.org/10.1103/PhysRevLett.92.140601} {\bibfield  {journal}
		{\bibinfo  {journal} {Phys. Rev. Lett.}\ }\textbf {\bibinfo {volume} {92}},\
		\bibinfo {pages} {140601} (\bibinfo {year} {2004})}\BibitemShut {NoStop}%
	\bibitem [{\citenamefont {Evans}\ and\ \citenamefont
		{Searles}(1994)}]{evans1994pre}%
	\BibitemOpen
	\bibfield  {author} {\bibinfo {author} {\bibfnamefont {D.~J.}\ \bibnamefont
			{Evans}}\ and\ \bibinfo {author} {\bibfnamefont {D.~J.}\ \bibnamefont
			{Searles}},\ }\bibfield  {title} {\bibinfo {title} {Equilibrium microstates
			which generate second law violating steady states},\ }\href
	{https://doi.org/10.1103/PhysRevE.50.1645} {\bibfield  {journal} {\bibinfo
			{journal} {Phys. Rev. E}\ }\textbf {\bibinfo {volume} {50}},\ \bibinfo
		{pages} {1645} (\bibinfo {year} {1994})}\BibitemShut {NoStop}%
	\bibitem [{\citenamefont {Khan}\ and\ \citenamefont
		{Sood}(2010)}]{khan2010out}%
	\BibitemOpen
	\bibfield  {author} {\bibinfo {author} {\bibfnamefont {M.}~\bibnamefont
			{Khan}}\ and\ \bibinfo {author} {\bibfnamefont {A.}~\bibnamefont {Sood}},\
	}\bibfield  {title} {\bibinfo {title} {Out-of-equilibrium microrheology using
			optical tweezers to probe directional viscoelastic properties under shear},\
	}\href@noop {} {\bibfield  {journal} {\bibinfo  {journal} {EPL (Europhysics
				Letters)}\ }\textbf {\bibinfo {volume} {92}},\ \bibinfo {pages} {48001}
		(\bibinfo {year} {2010})}\BibitemShut {NoStop}%
	\bibitem [{\citenamefont {Kumar}\ \emph {et~al.}(2011)\citenamefont {Kumar},
		\citenamefont {Ramaswamy},\ and\ \citenamefont {Sood}}]{kink_nitin_prl_2011}%
	\BibitemOpen
	\bibfield  {author} {\bibinfo {author} {\bibfnamefont {N.}~\bibnamefont
			{Kumar}}, \bibinfo {author} {\bibfnamefont {S.}~\bibnamefont {Ramaswamy}},\
		and\ \bibinfo {author} {\bibfnamefont {A.~K.}\ \bibnamefont {Sood}},\
	}\bibfield  {title} {\bibinfo {title} {Symmetry properties of the
			large-deviation function of the velocity of a self-propelled polar
			particle},\ }\href {https://doi.org/10.1103/PhysRevLett.106.118001}
	{\bibfield  {journal} {\bibinfo  {journal} {Phys. Rev. Lett.}\ }\textbf
		{\bibinfo {volume} {106}},\ \bibinfo {pages} {118001} (\bibinfo {year}
		{2011})}\BibitemShut {NoStop}%
	\bibitem [{\citenamefont {Kumar}\ \emph {et~al.}(2015)\citenamefont {Kumar},
		\citenamefont {Soni}, \citenamefont {Ramaswamy},\ and\ \citenamefont
		{Sood}}]{nk_hs_sr_as2014}%
	\BibitemOpen
	\bibfield  {author} {\bibinfo {author} {\bibfnamefont {N.}~\bibnamefont
			{Kumar}}, \bibinfo {author} {\bibfnamefont {H.}~\bibnamefont {Soni}},
		\bibinfo {author} {\bibfnamefont {S.}~\bibnamefont {Ramaswamy}},\ and\
		\bibinfo {author} {\bibfnamefont {A.~K.}\ \bibnamefont {Sood}},\ }\bibfield
	{title} {\bibinfo {title} {Anisotropic isometric fluctuation relations in
			experiment and theory on a self-propelled rod},\ }\href@noop {} {\bibfield
		{journal} {\bibinfo  {journal} {Physical Review E}\ }\textbf {\bibinfo
			{volume} {91}},\ \bibinfo {pages} {030102} (\bibinfo {year}
		{2015})}\BibitemShut {NoStop}%
	\bibitem [{\citenamefont {Aron}\ \emph {et~al.}(2016)\citenamefont {Aron},
		\citenamefont {Barci}, \citenamefont {Cugliandolo}, \citenamefont {Arenas},\
		and\ \citenamefont {Lozano}}]{aron2016dynamical}%
	\BibitemOpen
	\bibfield  {author} {\bibinfo {author} {\bibfnamefont {C.}~\bibnamefont
			{Aron}}, \bibinfo {author} {\bibfnamefont {D.~G.}\ \bibnamefont {Barci}},
		\bibinfo {author} {\bibfnamefont {L.~F.}\ \bibnamefont {Cugliandolo}},
		\bibinfo {author} {\bibfnamefont {Z.~G.}\ \bibnamefont {Arenas}},\ and\
		\bibinfo {author} {\bibfnamefont {G.~S.}\ \bibnamefont {Lozano}},\ }\bibfield
	{title} {\bibinfo {title} {Dynamical symmetries of markov processes with
			multiplicative white noise},\ }\href@noop {} {\bibfield  {journal} {\bibinfo
			{journal} {Journal of Statistical Mechanics: Theory and Experiment}\ }\textbf
		{\bibinfo {volume} {2016}},\ \bibinfo {pages} {053207} (\bibinfo {year}
		{2016})}\BibitemShut {NoStop}%
	\bibitem [{\citenamefont {Dengler}(2015)}]{dengler2015}%
	\BibitemOpen
	\bibfield  {author} {\bibinfo {author} {\bibfnamefont {R.}~\bibnamefont
			{Dengler}},\ }\bibfield  {title} {\bibinfo {title} {Another derivation of
			generalized langevin equations},\ }\href@noop {} {\bibfield  {journal}
		{\bibinfo  {journal} {arXiv preprint arXiv:1506.02650}\ } (\bibinfo {year}
		{2015})}\BibitemShut {NoStop}%
	\bibitem [{\citenamefont {Mazenko}(2008)}]{mazenkobook}%
	\BibitemOpen
	\bibfield  {author} {\bibinfo {author} {\bibfnamefont {G.~F.}\ \bibnamefont
			{Mazenko}},\ }\href@noop {} {\emph {\bibinfo {title} {Nonequilibrium
				statistical mechanics}}}\ (\bibinfo  {publisher} {John Wiley \& Sons},\
	\bibinfo {year} {2008})\BibitemShut {NoStop}%
	\bibitem [{\citenamefont {Lau}\ and\ \citenamefont
		{Lubensky}(2007)}]{lau2007state}%
	\BibitemOpen
	\bibfield  {author} {\bibinfo {author} {\bibfnamefont {A.~W.}\ \bibnamefont
			{Lau}}\ and\ \bibinfo {author} {\bibfnamefont {T.~C.}\ \bibnamefont
			{Lubensky}},\ }\bibfield  {title} {\bibinfo {title} {State-dependent
			diffusion: Thermodynamic consistency and its path integral formulation},\
	}\href@noop {} {\bibfield  {journal} {\bibinfo  {journal} {Physical Review
				E}\ }\textbf {\bibinfo {volume} {76}},\ \bibinfo {pages} {011123} (\bibinfo
		{year} {2007})}\BibitemShut {NoStop}%
	\bibitem [{\citenamefont {Cugliandolo}\ and\ \citenamefont
		{Lecomte}(2017)}]{cugliandolo2017rules}%
	\BibitemOpen
	\bibfield  {author} {\bibinfo {author} {\bibfnamefont {L.~F.}\ \bibnamefont
			{Cugliandolo}}\ and\ \bibinfo {author} {\bibfnamefont {V.}~\bibnamefont
			{Lecomte}},\ }\bibfield  {title} {\bibinfo {title} {Rules of calculus in the
			path integral representation of white noise langevin equations: the
			onsager--machlup approach},\ }\href@noop {} {\bibfield  {journal} {\bibinfo
			{journal} {Journal of Physics A: Mathematical and Theoretical}\ }\textbf
		{\bibinfo {volume} {50}},\ \bibinfo {pages} {345001} (\bibinfo {year}
		{2017})}\BibitemShut {NoStop}%
	\bibitem [{\citenamefont {Chaikin}\ and\ \citenamefont
		{Lubensky}(1995)}]{Lubensky_book}%
	\BibitemOpen
	\bibfield  {author} {\bibinfo {author} {\bibfnamefont {P.~M.}\ \bibnamefont
			{Chaikin}}\ and\ \bibinfo {author} {\bibfnamefont {T.~C.}\ \bibnamefont
			{Lubensky}},\ }\href {http://dx.doi.org/10.1017/CBO9780511813467} {\emph
		{\bibinfo {title} {Principles of Condensed Matter Physics}}}\ (\bibinfo
	{publisher} {Cambridge University Press},\ \bibinfo {year} {1995})\ \bibinfo
	{note} {cambridge Books Online}\BibitemShut {NoStop}%
	\bibitem [{\citenamefont {Onsager}(1931)}]{Onsager1931}%
	\BibitemOpen
	\bibfield  {author} {\bibinfo {author} {\bibfnamefont {L.}~\bibnamefont
			{Onsager}},\ }\bibfield  {title} {\bibinfo {title} {Reciprocal relations in
			irreversible processes. i.},\ }\href {https://doi.org/10.1103/PhysRev.37.405}
	{\bibfield  {journal} {\bibinfo  {journal} {Phys. Rev.}\ }\textbf {\bibinfo
			{volume} {37}},\ \bibinfo {pages} {405} (\bibinfo {year} {1931})}\BibitemShut
	{NoStop}%
	\bibitem [{\citenamefont {Gard}(1988)}]{tcgard}%
	\BibitemOpen
	\bibfield  {author} {\bibinfo {author} {\bibfnamefont {T.~C.}\ \bibnamefont
			{Gard}},\ }\href@noop {} {\emph {\bibinfo {title} {Introduction to stochastic
				differential equations}}}\ (\bibinfo  {publisher} {New York : M. Dekker},\
	\bibinfo {year} {1988})\BibitemShut {NoStop}%
	\bibitem [{\citenamefont {Sevick}\ \emph {et~al.}(2008)\citenamefont {Sevick},
		\citenamefont {Prabhakar}, \citenamefont {Williams},\ and\ \citenamefont
		{Searles}}]{ft_evan_review}%
	\BibitemOpen
	\bibfield  {author} {\bibinfo {author} {\bibfnamefont {E.}~\bibnamefont
			{Sevick}}, \bibinfo {author} {\bibfnamefont {R.}~\bibnamefont {Prabhakar}},
		\bibinfo {author} {\bibfnamefont {S.~R.}\ \bibnamefont {Williams}},\ and\
		\bibinfo {author} {\bibfnamefont {D.~J.}\ \bibnamefont {Searles}},\
	}\bibfield  {title} {\bibinfo {title} {Fluctuation theorems},\ }\href
	{https://doi.org/10.1146/annurev.physchem.58.032806.104555} {\bibfield
		{journal} {\bibinfo  {journal} {Annual Review of Physical Chemistry}\
		}\textbf {\bibinfo {volume} {59}},\ \bibinfo {pages} {603} (\bibinfo {year}
		{2008})},\ \bibinfo {note} {pMID: 18393680}\BibitemShut {NoStop}%
	\bibitem [{\citenamefont {Reid}\ \emph {et~al.}(2004)\citenamefont {Reid},
		\citenamefont {Carberry}, \citenamefont {Wang}, \citenamefont {Sevick},
		\citenamefont {Evans},\ and\ \citenamefont
		{Searles}}]{reid2004reversibility}%
	\BibitemOpen
	\bibfield  {author} {\bibinfo {author} {\bibfnamefont {J.}~\bibnamefont
			{Reid}}, \bibinfo {author} {\bibfnamefont {D.}~\bibnamefont {Carberry}},
		\bibinfo {author} {\bibfnamefont {G.}~\bibnamefont {Wang}}, \bibinfo {author}
		{\bibfnamefont {E.~M.}\ \bibnamefont {Sevick}}, \bibinfo {author}
		{\bibfnamefont {D.~J.}\ \bibnamefont {Evans}},\ and\ \bibinfo {author}
		{\bibfnamefont {D.~J.}\ \bibnamefont {Searles}},\ }\bibfield  {title}
	{\bibinfo {title} {Reversibility in nonequilibrium trajectories of an
			optically trapped particle},\ }\href@noop {} {\bibfield  {journal} {\bibinfo
			{journal} {Physical Review E}\ }\textbf {\bibinfo {volume} {70}},\ \bibinfo
		{pages} {016111} (\bibinfo {year} {2004})}\BibitemShut {NoStop}%
	\bibitem [{\citenamefont {Marchetti}\ \emph {et~al.}(2013)\citenamefont
		{Marchetti}, \citenamefont {Joanny}, \citenamefont {Ramaswamy}, \citenamefont
		{Liverpool}, \citenamefont {Prost}, \citenamefont {Rao},\ and\ \citenamefont
		{Simha}}]{RevModPhys.85.1143}%
	\BibitemOpen
	\bibfield  {author} {\bibinfo {author} {\bibfnamefont {M.~C.}\ \bibnamefont
			{Marchetti}}, \bibinfo {author} {\bibfnamefont {J.~F.}\ \bibnamefont
			{Joanny}}, \bibinfo {author} {\bibfnamefont {S.}~\bibnamefont {Ramaswamy}},
		\bibinfo {author} {\bibfnamefont {T.~B.}\ \bibnamefont {Liverpool}}, \bibinfo
		{author} {\bibfnamefont {J.}~\bibnamefont {Prost}}, \bibinfo {author}
		{\bibfnamefont {M.}~\bibnamefont {Rao}},\ and\ \bibinfo {author}
		{\bibfnamefont {R.~A.}\ \bibnamefont {Simha}},\ }\bibfield  {title} {\bibinfo
		{title} {Hydrodynamics of soft active matter},\ }\href
	{https://doi.org/10.1103/RevModPhys.85.1143} {\bibfield  {journal} {\bibinfo
			{journal} {Rev. Mod. Phys.}\ }\textbf {\bibinfo {volume} {85}},\ \bibinfo
		{pages} {1143} (\bibinfo {year} {2013})}\BibitemShut {NoStop}%
	\bibitem [{\citenamefont {Bricard}\ \emph {et~al.}(2013)\citenamefont
		{Bricard}, \citenamefont {Caussin}, \citenamefont {Desreumaux}, \citenamefont
		{Dauchot},\ and\ \citenamefont {Bartolo}}]{Bricard2013}%
	\BibitemOpen
	\bibfield  {author} {\bibinfo {author} {\bibfnamefont {A.}~\bibnamefont
			{Bricard}}, \bibinfo {author} {\bibfnamefont {J.-B.}\ \bibnamefont
			{Caussin}}, \bibinfo {author} {\bibfnamefont {N.}~\bibnamefont {Desreumaux}},
		\bibinfo {author} {\bibfnamefont {O.}~\bibnamefont {Dauchot}},\ and\ \bibinfo
		{author} {\bibfnamefont {D.}~\bibnamefont {Bartolo}},\ }\bibfield  {title}
	{\bibinfo {title} {Emergence of macroscopic directed motion in populations of
			motile colloids},\ }\href {http://dx.doi.org/10.1038/nature12673} {\bibfield
		{journal} {\bibinfo  {journal} {Nature}\ }\textbf {\bibinfo {volume} {503}},\
		\bibinfo {pages} {95} (\bibinfo {year} {2013})}\BibitemShut {NoStop}%
	\bibitem [{\citenamefont {Kumar}\ \emph {et~al.}(2014)\citenamefont {Kumar},
		\citenamefont {Soni}, \citenamefont {Ramaswamy},\ and\ \citenamefont
		{Sood}}]{hs_nk}%
	\BibitemOpen
	\bibfield  {author} {\bibinfo {author} {\bibfnamefont {N.}~\bibnamefont
			{Kumar}}, \bibinfo {author} {\bibfnamefont {H.}~\bibnamefont {Soni}},
		\bibinfo {author} {\bibfnamefont {S.}~\bibnamefont {Ramaswamy}},\ and\
		\bibinfo {author} {\bibfnamefont {A.~K.}\ \bibnamefont {Sood}},\ }\bibfield
	{title} {\bibinfo {title} {Flocking at a distance in active granular
			matter},\ }\href {http://dx.doi.org/10.1038/ncomms5688} {\bibfield  {journal}
		{\bibinfo  {journal} {Nat Commun}\ }\textbf {\bibinfo {volume} {5}} (\bibinfo
		{year} {2014})}\BibitemShut {NoStop}%
	\bibitem [{\citenamefont {Weber}\ \emph {et~al.}(2013)\citenamefont {Weber},
		\citenamefont {Hanke}, \citenamefont {Deseigne}, \citenamefont {L\'eonard},
		\citenamefont {Dauchot}, \citenamefont {Frey},\ and\ \citenamefont
		{Chat\'e}}]{PhysRevLett.110.208001}%
	\BibitemOpen
	\bibfield  {author} {\bibinfo {author} {\bibfnamefont {C.~A.}\ \bibnamefont
			{Weber}}, \bibinfo {author} {\bibfnamefont {T.}~\bibnamefont {Hanke}},
		\bibinfo {author} {\bibfnamefont {J.}~\bibnamefont {Deseigne}}, \bibinfo
		{author} {\bibfnamefont {S.}~\bibnamefont {L\'eonard}}, \bibinfo {author}
		{\bibfnamefont {O.}~\bibnamefont {Dauchot}}, \bibinfo {author} {\bibfnamefont
			{E.}~\bibnamefont {Frey}},\ and\ \bibinfo {author} {\bibfnamefont
			{H.}~\bibnamefont {Chat\'e}},\ }\bibfield  {title} {\bibinfo {title}
		{Long-range ordering of vibrated polar disks},\ }\href
	{https://doi.org/10.1103/PhysRevLett.110.208001} {\bibfield  {journal}
		{\bibinfo  {journal} {Phys. Rev. Lett.}\ }\textbf {\bibinfo {volume} {110}},\
		\bibinfo {pages} {208001} (\bibinfo {year} {2013})}\BibitemShut {NoStop}%
	\bibitem [{\citenamefont {Chatterjee}\ \emph {et~al.}(2021)\citenamefont
		{Chatterjee}, \citenamefont {Rana}, \citenamefont {Simha}, \citenamefont
		{Perlekar},\ and\ \citenamefont {Ramaswamy}}]{chatterjee2021inertia}%
	\BibitemOpen
	\bibfield  {author} {\bibinfo {author} {\bibfnamefont {R.}~\bibnamefont
			{Chatterjee}}, \bibinfo {author} {\bibfnamefont {N.}~\bibnamefont {Rana}},
		\bibinfo {author} {\bibfnamefont {R.~A.}\ \bibnamefont {Simha}}, \bibinfo
		{author} {\bibfnamefont {P.}~\bibnamefont {Perlekar}},\ and\ \bibinfo
		{author} {\bibfnamefont {S.}~\bibnamefont {Ramaswamy}},\ }\bibfield  {title}
	{\bibinfo {title} {Inertia drives a flocking phase transition in viscous
			active fluids},\ }\href@noop {} {\bibfield  {journal} {\bibinfo  {journal}
			{Physical Review X}\ }\textbf {\bibinfo {volume} {11}},\ \bibinfo {pages}
		{031063} (\bibinfo {year} {2021})}\BibitemShut {NoStop}%
	\bibitem [{\citenamefont {Doi}\ \emph {et~al.}(1988)\citenamefont {Doi},
		\citenamefont {Edwards},\ and\ \citenamefont {Edwards}}]{doibookpolymer}%
	\BibitemOpen
	\bibfield  {author} {\bibinfo {author} {\bibfnamefont {M.}~\bibnamefont
			{Doi}}, \bibinfo {author} {\bibfnamefont {S.~F.}\ \bibnamefont {Edwards}},\
		and\ \bibinfo {author} {\bibfnamefont {S.~F.}\ \bibnamefont {Edwards}},\
	}\href@noop {} {\emph {\bibinfo {title} {The theory of polymer dynamics}}},\
	Vol.~\bibinfo {volume} {73}\ (\bibinfo  {publisher} {oxford university
		press},\ \bibinfo {year} {1988})\BibitemShut {NoStop}%
\end{thebibliography}
\end{document}